\documentclass[prd,nofootinbib,preprint,superscriptaddress,showpacs]{revtex4}

\usepackage{graphicx, epsfig, color}

\def\eslt{\not\!\!{E_T}}
\def\eslt{E_T^{\rm miss}}
\def\ba{\begin{array}}
\def\ea{\end{array}}
\def\to{\rightarrow}

\def\bi{\begin{itemize}}
\def\ei{\end{itemize}}
\def\te{\tilde e}

\def\tu{\tilde u}

\def\tb{\tilde b}

\def\td{\tilde d}

\def\tst{\tilde t}
\def\ttau{\tilde \tau}

\def\tg{\tilde g}
\def\tnu{\tilde\nu}

\def\tq{\tilde q}
\def\tw{\widetilde W}
\def\tz{\widetilde Z}

\def\be{\begin{equation}}  
\def\ee{\end{equation}}  
\newcommand\PRD[3]{{Phys.\ Rev.\ D }{\bf #1} (#2) #3}

\newcommand\PRL[3]{{Phys.\ Rev.\ Lett.\ }{\bf #1} (#2) #3}
\newcommand\plb[3]{{Phys.\ Lett.\ }{\bf B #1} (#2) #3}
\newcommand\jhep[3]{{J.\ High Energy Phys.\ }{\bf #1} (#2) #3}
\newcommand\app[3]{{Astropart.\ Phys.\ }{\bf #1} (#2) #3}

\newcommand\npb[3]{{Nucl.\ Phys.\ }{\bf B #1} (#2) #3}
\newcommand\epjc[3]{{Eur.\ Phys.\ J.\ }{\bf C #1} (#2) #3}
\newcommand\ptp[3]{{Prog.\ Theor.\ Phys.\ }{\bf #1} (#2) #3}
\newcommand{\hepph}[1]{hep-ph/#1}

\begin{document}
%\preprint{\hbox{IFT-P.XXX/2008}}
\preprint{\hbox{UH-511-1122-08}}

\title{Heavy-flavour tagging and the supersymmetry reach of the CERN Large
Hadron Collider}

\author{R.~H.~K.~Kadala}
\email{kadala@phys.hawaii.edu}
\affiliation{College of Natural Sciences, Hawaii Pacific University,
  Kaneohe, HI 96744.}
\affiliation{Department of Physics and Astronomy, University of Hawaii,
  Honolulu, HI 96822.}
\author{P.~G.~Mercadante}
\email{mercadan@fnal.gov}
\affiliation{Instituto de F\'{\i}sica Te\'orica, 
Universidade Estadual Paulista, S\~ao Paulo -- SP, Brazil.}
\author{J.~K.~Mizukoshi}
\email{mizuka@ufabc.edu.br}
\affiliation{Centro de Ci\^encias Naturais e Humanas,
Universidade Federal do ABC, Santo Andr\'e -- SP, Brazil.}
\author{Xerxes Tata}
\email{tata@phys.hawaii.edu}
\affiliation{Department of Physics and Astronomy, University of Hawaii,
  Honolulu, HI 96822.}

\begin{abstract}
The branching fraction for the decays of gluinos to third
generation quarks is expected to be enhanced in classes of
supersymmetric models where either third generation squarks are lighter
than other squarks, or in mixed-higgsino dark matter models constructed
to be in concordance with the measured density of cold dark matter.  In
such scenarios, gluino production events at the CERN Large Hadron
Collider should be rich in top and bottom quark jets. Requiring $b$-jets
in addition to $\eslt$ should, therefore, enhance the supersymmetry
signal relative to Standard Model backgrounds from $V$ + jet, $VV$ and
QCD backgrounds ($V=W, Z$). We quantify the increase in the
supersymmetry reach of the LHC from $b$-tagging in a variety of
well-motivated models of supersymmetry. We also explore ``top-tagging''
at the LHC. We find that while the efficiency for this turns out to be
too low to give an increase in reach beyond that obtained via
$b$-tagging, top-tagging can indeed provide a confirmatory signal if
gluinos are not too heavy. We also examine $c$-jet tagging but find that
it is not useful at the LHC. Finally, we explore the prospects for
detecting the direct production of third generation squarks in models
with an inverted squark mass hierarchy. This is signalled by $b$-jets +
$\eslt$ events harder than in the Standard Model, but softer than those
from the production of gluinos and heavier squarks.  We find that while
these events can be readily separated from SM background (for third
generation squark masses $\sim 300-500$~GeV), the contamination from the
much heavier gluinos and squarks remains formidable if these are also
accessible.
\end{abstract}


\maketitle 

%%%%%%%%%%%%%%%%%%%%%%%%%%%%%%%%%%%%%%%%%%%%%%%%%%%%%%%%%%%%%%%%%%%%%%
\section{Introduction} \label{intro}

Weak scale supersymmetry (SUSY) \cite{wss} is the best motivated and
most carefully studied extension of the Standard Model (SM).  The
hypothesis of TeV scale super-partners of SM particles simultaneously
stabilizes the gauge hierarchy, accounts for gauge coupling unification,
and naturally accommodates the measured relic density in the minimal
extension of the SM.  This is especially exciting because squarks and
gluinos will be produced at observable rates at the Large Hadron
Collider (LHC) if their masses are smaller than 2-3~TeV
\cite{bcpt,atlas,cms,other,update}. Supersymmetric models generically allow
renormalizable baryon- and lepton-number violating operators that lead
to proton decay at the typical weak interaction rate, and would be
strongly excluded unless there is an additional symmetry that forbids
these interactions.  Assuming that $R$-parity serves this purpose, heavy
superpartners must decay into lighter sparticles until the decay
terminates in the lightest supersymmetric particle (LSP), which must be
stable. Cosmological considerations require that the LSP is electrically
and colour neutral so that it escapes the experimental apparatus without
significant deposition of energy. Then, super-partner production at colliders
is generically signalled by multi-jet plus multi-lepton events with
large amounts of $\eslt$ carried off by the escaping LSPs. We will
assume that the lightest neutralino is the LSP as is the case in many
models.

Remarkably, weak scale SUSY models with a stable neutralino LSP
naturally lead to the right magnitude for the measured relic density of
thermally produced cold dark matter \cite{wmap},
\be
\Omega_{\rm CDM}h^2=0.111^{+0.011}_{-0.015} \ \ (2\sigma)\;,
\label{wmap1}
\ee
if
superpartner masses are $\sim 100$~GeV. Assuming thermal production and
standard Big Bang cosmology, the upper limit from (\ref{wmap1}) 
provides a stringent constraint on any theory with stable weakly
interacting particles, in particular on weak scale SUSY theories. Since
the dark matter may well consist of several components, the contribution
from any single component may well not saturate the observed value, so
that strictly speaking the relic density measurement serves as an upper
bound, 
\be
\Omega_{\tz_1}h^2 < 0.12\;,
\label{wmap}
\ee
on the relic density of neutralinos, or for that matter, on the density
of any other stable particle.

Direct searches for charged sparticles at LEP~2 have
resulted in lower limits of about 100~GeV on chargino and selectron
masses, and slightly lower on the masses of smuons and staus \cite{lep2}. Since
neutralinos can annihilate via $t$-channel sfermion exchange, the
measured value of the relic density, on the other hand, favours
sfermions lighter than about 100~GeV, resulting in some tension with the
LEP~2 bounds. In many constrained models where all sparticle masses and
couplings are fixed by just a few parameters, such light sparticles
often also lead to measurable deviations in {\it other} observables, and
hence are disfavoured. If the SUSY mass scale is raised to avoid
these constraints, the annihilation cross-section which is proportional
to $\frac{1}{M_{\rm SUSY}^2}$ is correspondingly reduced, and the
neutralino relic density turns out to be too large.  One way to fix this is
by invoking non-thermal relics or non-standard cosmology to dilute the
relic density. However, it seems much more economical to invoke
SUSY mechanisms that enhance the neutralino annihilation rate to bring their
thermal relic density in line with (\ref{wmap1}).

The primary reason for the low neutralino annihilation rate lies in the
fact that the LSP is dominantly a bino in many models with assumed
gaugino mass unification, where the bino and wino masses are related by
$M_1 \simeq {1\over 2} M_2$. The annihilation of bino pairs to gauge
bosons is forbidden because $SU(2)\times U(1)$ precludes the couplings
of binos to the gaugino-gauge boson system, while annihilation to
fermions may be suppressed by large sfermion masses and the relatively
small hypercharge coupling. Finally, annihilation to Higgs boson pairs
is suppressed by the (usually large) higgsino mass, as well as by the
relatively small hypercharge gauge coupling.  This then suggests several
ways in which the neutralino annihilation rate may be enhanced to bring
their thermal relic density in accord with (\ref{wmap1}).
\begin{itemize}
\item We can arrange the mass of a charged or coloured sparticle to be
  close to that of the LSP. Since these coloured/charged sparticles can
  annihilate efficiently, interactions between them and the neutralino
  which maintain thermal equilibrium will necessarily also reduce the
  neutralino relic density \cite{GS}. Within the mSUGRA model, the
  co-annihilating sparticle is usually either the scalar tau \cite{stau}
  or the scalar top \cite{stop}, but different  choices are possible in
  other models.
  
\item We can arrange $2m_{\tz_1} \simeq m_A \simeq m_H$, so that
  neutralino annihilation is resonantly enhanced through $s$-channel
  heavy Higgs boson exchange \cite{Afunnel}. The large widths of $A$ and
  $H$ together with the thermal motion of the LSPs in the early universe
  then enhances the annihilation cross section over a considerable range
  of parameters. Within the mSUGRA model, this is possible only if
  $\tan\beta$ is very large. However, in models with non-universal Higgs
  mass (NUHM) parameters, where the Higgs scalar mass parameters do not
  unify with matter scalar parameters as in mSUGRA \cite{NUHMgen,NUHM},
  agreement with (\ref{wmap1}) may be obtained via resonant $A/H$
  annihilation for any value of $\tan\beta$.  We mention that
%arXiv:0707.0618 (2007) [hep-ph]; 
  resonantly-enhanced annihilation may also occur via $h$ exchange,
  albeit for a much smaller range of parameters \cite{hfunnel}.

\item It is also possible to obtain an enhanced neutralino annihilation
  rate if the light top squark, $\tst_1$, is relatively light so that
  neutralinos efficiently annihilate via $\tz_1\tz_1 \to t\bar{t}$
  \cite{compressed}, or in
  NUHM models via $\tz_1\tz_1 \to u\bar{u}$ or $c\bar{c}$, via
  $t$-channel top- or right-squark exchanges, respectively \cite{NUHM}. 
\end{itemize}

Instead of adjusting sparticle masses, we can also adjust the
composition of the neutralino. More specifically,
\begin{itemize}
\item We can increase the higgsino content of the neutralino so that its
  coupling to electroweak gauge bosons is increased, leading to
  mixed higgsino dark matter (MHDM). Within the mSUGRA framework, we 
  can only do so in the so-called hyperbolic branch/focus point
  (HB/FP) region where $m_0$ takes on multi-TeV values \cite{focus}, but
  in NUHM models this is possible for {\it all} values of $m_0$
  \cite{NUHM}. The higgsino content may also be increased by relaxing
  the assumed high scale universality between gaugino masses. The
  usually assumed universality of gaugino masses follows if the
  auxiliary field that breaks supersymmetry does not break the
  underlying grand unification symmetry; if this is not the case,
  non-universal gaugino masses can result. It has been shown that if the
  GUT scale gluino mass is smaller than the other gaugino masses,
  $m_{H_u}^2$ does not run as negative as usual, yielding a smaller
  value of $\mu^2$, resulting in an increased higgsino content of $\tz_1$
  \cite{lm3dm}.  This has been dubbed as low $M_3$ dark matter
  (LM3DM). Very recently \cite{hm2dm} it has been pointed out that
  increasing the GUT scale wino mass parameter from its unified value
  also results in a low value of $|\mu|$, resulting in consistency with
  (\ref{wmap1}) via MHDM.

\item Finally, depending on the gauge transformation property of the
  SUSY breaking auxiliary field, it may also be possible to enhance the
  wino content of the neutralino leading to mixed wino dark matter
  (MWDM) \cite{mwdm}. This requires that the weak scale values of bino
  and wino masses to be approximately equal. If instead these are
  roughly equal in magnitude but differ in sign, bino-wino mixing is
  suppressed, but agreement with the observed relic density is possible
  via bino-wino co-annihilation (BWCA) \cite{bwca}.

\end{itemize}

These various mechanisms result in characteristic modifications of
supersymmetry signals at the LHC, at the proposed linear
electron-positron collider (LC) or at experiments for direct and
indirect detection of the relic neutralinos in our galactic halo
\cite{rev}. Of interest to us here is the potential for an enhanced rate
for bottom quark production in SUSY events that occurs for MHDM, as
exemplified by (but not limited to) the HB/FP region of the mSUGRA model
\cite{DDDR,MMT}, or models where third generation squarks are
significantly lighter than other squarks as, for instance, in the stop
co-annihilation region of mSUGRA, in so-called inverted hierarchy models
where third generation sfermions are much lighter than those of the
first two generations \cite{bagger,imh}, or in the framework suggested
in Ref.~\cite{compressed}.

We have multiple goals for this study. First, we follow up an earlier
investigation \cite{MMT} by three of us where we showed that using
$b$-jet tagging techniques that are available at the LHC, the SUSY reach
may be enhanced by as much as 20\% for parameters in the HB/FP of the
mSUGRA model. Toward this end, we examine the reach of the LHC with and
without $b$-jet tagging, in several models motivated by the relic
density measurement just discussed\footnote{More precisely, when we
refer to models satisfying (\ref{wmap1}) below, we require the
neutralino relic density to be close to its measured central value so
that the neutralino is the dominant DM component, but strictly
satisfying (\ref{wmap}). The impact on the LHC reach from $b$-jet
tagging is insensitive to the precise value of $\Omega_{\tz_1}h^2$.}  as
well as by other considerations. We find that the reach is increased by
different amounts, and that sometimes requiring $b$-tagging even reduces
the reach.  One aim of this study is to precisely delineate the
circumstances under which $b$-jet tagging will significantly enhance the
LHC reach. Second, since SUSY events are frequently also be enriched in
$t$-jets, we examine prospects for top jet tagging in SUSY events at the
LHC. Third, motivated by the fact that $c$-tagging has been suggested
\cite{sender} as a way to enhance the $t$-squark SUSY signal at the
Femilab Tevatron, we also examine whether tagging charm jets may serve
to increase the SUSY reach of the LHC. Finally, since third generation
squarks are expected to be lighter than other squarks in many models, we
explore the prospects for using $b$-tagging to isolate signals from
their direct production from both SM backgrounds as well as from other
sparticle production processes. 

The rest of this paper is organized as follows. In Sec.~\ref{models} we
introduce the various models that offer the potential for an enhanced
$b$-jet signal, and discuss the parameter space for each of these
models. In Sec.~\ref{sim}, we discuss our event simulation using ISAJET
\cite{isajet}, and how we use this to model the LHC environment. In
Sec.~\ref{sec:btag} we examine a large set of selection cuts that may be
used to optimize the SUSY signal, and design a set of cuts that we
believe should work for a wide class of models over their entire
parameter range: we then use these to obtain projections for how $b$-jet
tagging would enhance the LHC reach for these models. We discuss the
prospects for top tagging in Sec.~\ref{sec:toptag}. In Sec.~\ref{third},
we describe our strategy for isolating the signal for the direct
production of third generation squarks from SM backgrounds as well as
from contamination from the production of gluinos and heavier squarks,
since the observation of such a signal would unequivocally point to
models with an inverted mass hierarchy.  Finally, we report on our
(negative) results for using charm-tagging to enhance the SUSY signal at
the LHC in Sec.~\ref{ctag}.  We summarize our findings in
Sec.~\ref{summary}.

\section{Models} \label{models}

In this section, we discuss several models in which we may expect third
generation fermions to be preferentially produced in SUSY models. We
begin with the familiar mSUGRA model, and work our way through
various other models motivated either by the relic density observation
discussed in Sec.~\ref{intro}, or by other considerations.

\subsection{The mSUGRA model} The mSUGRA model \cite{msugra}, whose
hallmark is the unification of soft SUSY breaking (SSB) parameters
renormalized at a scale $Q\simeq M_{\rm GUT}$ to $M_{\rm Planck}$, has served as
the paradigm for many phenomenological analyses of SUSY. Assuming that
the radiative electroweak symmetry breaking mechanism is operative 
\cite{rewsb}, the
observed value of $M_Z^2$ can be used to fix $\mu^2$, and the framework
is completely specified by the well known parameter set,
\be
m_0, m_{1/2}, \tan\beta, A_0 \ \ {\rm and} \  {\rm sign}(\mu)\;.
\ee
Typically, the weak scale value of $|\mu|$ is similar in magnitude to
$m_{\tg}$, and the bino is the LSP. However, for any chosen value of $m_{1/2}$,
the requirement that electroweak symmetry be correctly broken imposes an
upper bound on $m_0$, since the value of $\mu^2$ becomes negative for
yet larger values of $m_0$. There is thus a contour in the $m_0-m_{1/2}$
plane where $\mu^2=0$. For values of $m_0$ just below this bound, $\mu^2
\ll m_{\tg}^2$ and can be comparable to the SSB bino mass parameter,
$M_1$, so that the lightest neutralino is a mixed bino-higgsino state
that can annihilate rapidly in the early universe, mainly via its higgsino
content. This is the celebrated HB/FP region of the mSUGRA model
\cite{focus}, one of the regions of mSUGRA parameter space where the
expected neutralino relic density is consistent with (\ref{wmap1})
\cite{wilczek}. For parameters in this region, squark masses are in the
multi-TeV range, and
the reach of the LHC is determined by final states from gluino pair
production: although the higgsino-like chargino may be light, the mass
difference $m_{\tw_1}-m_{\tz_1}$ is small so that leptons from its decays
are too soft to increase the reach beyond that obtained via the $\eslt$
signal from gluino pair production \cite{howiefocus}. Since the LSP couples
preferentially to the third family via its higgsino component, cascade
decays of the gluino to third generation fermions tend to be
enhanced. As a result, the requirement of a $b$-tagged jet in SUSY events
reduces SM backgrounds and enhances the LHC reach by 15--20\% beyond the
reach via the inclusive $\eslt$ channel in the HB/FP region of the
mSUGRA model \cite{MMT}.

We should also mention that the $b$-jet multiplicity may also be
enhanced in the mSUGRA model if third generation squarks happen to be
light, either because of large bottom quark Yukawa couplings when $\tan\beta$
is large, or because the $A_t$ parameter happens to be ``just right'' so
that $m_{\tst_1}\ll m_{\tq}$, and $\tst_1$ mainly decays via $\tst_1 \to
b\tw_1 \ {\rm and} \  t\tz_1$, or $\tst_1 \to bW\tz_1$.

\subsection{Inverted mass hierarchy models} 

The evidence for neutrino
oscillations \cite{neutrino} and its interpretation in terms of neutrino
masses provides strong motivation for considering $SO(10)$ SUSY grand
unified theories (GUTS) \cite{soten}. Each generation of matter (including the
sterile neutrino) can be unified into a single ${\bf 16}$ dimensional
representation of $SO(10)$ while the Higgs superfields $\hat{H}_u$ and
$\hat{H_d}$ are both contained in a single ${\bf 10}$ dimensional
representation, allowing for the unification of both gauge (and
separately) Yukawa couplings. $SO(10)$ may either be directly broken to
the SM gauge group, or by a two step process via an intermediate stage
of $SU(5)$ unification. The spontaneous breakdown of $SO(10)$ with the
concomitant reduction of rank leaves an imprint on the SSB masses which
is captured by one additional parameter $M_D^2$ with a weak scale
magnitude but which can take either sign \cite{dterm}. The model
is then completely specified by the parameter set,
\be
m_{16}, m_{10}, m_{1/2}, M_D^2, \tan\beta, A_0 \ {\rm and} \ {\rm sign}(\mu)\;.
\ee
where we have assumed a common SSB mass parameter $m_{16}$ and a
different parameter $m_{10}$ for matter and Higgs fields in the ${\bf
16}$ and ${\bf 10}$ dimensional representations, respectively. The GUT
scale SSB masses for MSSM fields then take the form \cite{dterm},
$$ m_Q^2 = m_E^2 = m_U^2 = m_{16}^2 + M_D^2\;,$$ 
\be m_D^2 = m_L^2 = m_{16}^2 - 3 M_D^2\;, \ee
$$ m_{N}^2 = m_{16}^2 + 5 M_{D}^2\;, $$
$$m_{H_{u,d}}^2 = m_{10}^2 \mp 2 M_D^2\;. $$   
Unification of Yukawa couplings is possible for very large values of
$\tan\beta$ \cite{raby,yukunif}.\footnote{These studies require only
approximate unification of third generation quark Yukawa couplings to
allow for threshold effects. It has been argued that {\it exact}
unification of these Yukawa couplings leads to a tension with
flavour-violation in the $B$ and $B_s$ meson systems unless sparticles
are significantly heavier that $\sim$1~TeV, or a more complicated
flavour structure is introduced into the SUSY-breaking sector
\cite{buras}. This tension is alleviated if small deviations from exact
Yukawa coupling unification are admitted \cite{altmann}.}

The $SO(10)$ framework that we have just introduced naturally allows a
phenomenologically interesting class of models in which the ordering of matter
sfermion masses is inverted with respect to the order for the corresponding
fermions \cite{bagger}. Specifically, in models with Yukawa coupling
unification, the choice
\be A_0^2 = 2m_{10}^2 = 4m_{16}^2 
\label{so10rel}
\ee 
for the SSB parameters serves to drive third generation sfermion mass
parameters to sub-TeV values, leaving first and second generation
scalars as heavy as 2--3~TeV. A positive value of $M_D^2 \lesssim
(m_{16}/3)^2$ is necessary to obtain radiative electroweak symmetry
breaking \cite{imh}. The multi-TeV values of first and second generation scalar
masses ameliorate the SUSY $CP$ and flavour problems without destroying
the SUSY resolution of the gauge hierarchy problem, since the fields
with substantial direct couplings to the Higgs sector (gauginos and
third generation scalars) have masses below the TeV scale.  Because
third generation sfermions are significantly lighter than their
first/second generation cousins, we may expect that SUSY events are
enriched in $b$- (and possibly $t$-) quark jets in this scenario.

\subsection{Non-Universal Higgs Mass Models} Within the mSUGRA model, if
$m_0^2=m_{H_u}^2({\rm GUT})$ is smaller than or comparable to $m_{1/2}^2$, 
$m_{H_u}^2$ runs to a large negative value at the weak scale. The
minimization condition for the (tree level) Higgs scalar potential which
reads
\be
\mu^2 =\frac{m_{H_d}^2-m_{H_u}^2\tan^2\beta}{\tan^2\beta-1} 
-\frac{M_Z^2}{2} \simeq -m_{H_u}^2 -\frac{M_Z^2}{2}
\label{lowmu}
\ee
(where the last approximation is valid for moderate to large values of
$\tan\beta$), then implies that 
$|\mu| \gg |M_{1,2}|$ so that the LSP is essentially
a bino, while the heavier -inos are mainly higgsino-like. A way
of avoiding this conclusion is to choose $m_{H_u}^2({\rm GUT})$ 
such that $m_{H_u}^2$ runs to small negative values at the weak
scale. Within the mSUGRA model, this can only be realized by choosing $m_0 \gg
m_{1/2}$ which gives us the well studied HB/FP region with MHDM
discussed above. 

A different way would be to relax the assumed universality
\cite{NUHMgen} between the matter scalar and Higgs boson SSB mass
parameters in what has been dubbed as non-universal Higgs mass (NUHM)
models, and adopt a large value for $m_{H_u}^2({\rm GUT})$. In order to
avoid unwanted flavour changing neutral currents, we maintain a
universal value $m_0$ for matter scalars. The GUT scale value of the SSB
down Higgs mass parameter may (may not) be equal to $m_{H_u}^2$ leading
to a one (two) parameter extension of the mSUGRA framework that we will
refer to as the NUHM1 (NUHM2) model \cite{NUHM}. The NUHM1 model is thus
completely specified by the mSUGRA parameter set together with
$m_{\phi}={\rm sign}(m_{H_{u,d}}^2)\sqrt{|m_{H_{u,d}}^2|}$, {\it i.e.}
by,
\be m_0, m_{\phi}, m_{1/2}, A_0, \tan\beta \ {\rm and} \ {\rm sign}(\mu)
\ \ ({\rm NUHM1})\;.  
\ee 
If $m_\phi$ is chosen to be sufficiently larger than
$m_0$, the parameter $m_{H_u}^2$ runs down to negative values but
remains small in magnitude so that we obtain MHDM {\it for any value of
$m_0$ and $m_{1/2}$}.\footnote{Of course, if $m_{\phi}$ is chosen to be
  too large then $m_{H_u}^2$ does not run to negative values and
  electroweak symmetry breaking is no longer obtained.}

Curiously, the NUHM1 model accommodates another possibility of getting
agreement with (\ref{wmap1}). If $m_{\phi} < 0$, $m_{H_u}^2$ and
$m_{H_d}^2$ both run to large, negative values at the weak scale so that 
\be
m_A^2 = m_{H_u}^2 + m_{H_d}^2 +2\mu^2 \simeq m_{H_d}^2-m_{H_u}^2 - M_Z^2
\ee
may be small enough for neutralinos to annihilate via the $A$ and $H$
resonances. Within the NUHM1 framework, the Higgs funnel thus occurs for
{\it all values of $\tan\beta$.} Since the Higgs bosons $A$ and $H$
with relatively small masses are expected
to be produced via cascade decays of gluinos and squarks, and since
these decay preferentially to third generation fermions,  we may once
again expect an enhancement of the $b$- and, perhaps also, $t$-jet
multiplicity. 

The NUHM2 model requires two more parameters than the mSUGRA framework
for its complete specification. While these may be taken to be the GUT
scale values of $m_{H_u}^2$ and $m_{H_d}^2$, it is customary and more
convenient to eliminate these in favour of $m_A$ and $\mu$, and work
with the hybrid parameter set,
\be 
m_0, m_{1/2}, m_A, \mu, A_0, \tan\beta  \ \ ({\rm NUHM2})\;.  
\ee 
This then allows us to adjust the higgsino content of charginos and
neutralinos at will, and furthermore allows as much freedom in the
(tree-level) Higgs sector as in the unconstrained MSSM.

\subsection{Low ${\bf |M_3|}$ Dark Matter Model} Instead of relaxing the
universality between scalar masses as in the NUHM model, we can also
relax the universality between the gaugino mass parameters. If we adjust
the GUT scale value of $M_1/M_2$ so that $M_1\simeq M_2$ at the weak
scale, we obtain mixed wino DM \cite{mwdm}. Since there is no principle
that forces $M_1/M_2$ to be positive, we can instead adjust this ratio
so that $M_1 \simeq -M_2$ at the weak scale. In this case the LSP
remains a bino with charged and neutral winos close in mass to it and
agreement with (\ref{wmap1}) is obtained via bino-wino co-annihilation
\cite{bwca}. Although collider signatures are indeed altered from mSUGRA
expectations, we do not expect any enrichment of $b$-jet multiplicity in
this case.

Although not immediately obvious, agreement with (\ref{wmap1}) is also
obtained if we maintain $M_1=M_2$ at $Q=M_{\rm GUT}$, but instead {\it
reduce} the value of $|M_3|$. Specifically, for smaller values of
$|M_3|$, the (top)-squark mass parameters and also $A_t^2$ are driven to
smaller values at the weak scale. These smaller values of top-squark
masses and of $A_t^2$, in turn, slow down the evolution of $m_{H_u}^2$
so that it runs to negative values more slowly than in the mSUGRA
model. As a result, the weak scale value of $m_{H_u}^2$ though negative,
has a smaller magnitude than in the mSUGRA case, so that the value of
$\mu^2$ is correspondingly reduced [see Eq.~(\ref{lowmu})] and the LSP
becomes MHDM \cite{lm3dm}. This is referred to as the low $|M_3|$ DM
(LM3DM) model, and the corresponding parameter space is given by,
\be
m_0, m_{1/2}, M_3, A_0, \tan\beta, {\rm sign}(\mu) \ \ ({\rm LM3DM})\;.
\ee
Here $m_{1/2}> 0$ denotes the GUT scale value of $M_1=M_2$, while $M_3$
(which is either positive or negative) denotes the corresponding value
of $M_3$ at the GUT scale. For $m_0 \sim m_{1/2} \lesssim 1$~TeV, the GUT
scale value of $|M_3|$ must be {\it reduced} from its mSUGRA value in
order to obtain MHDM as discussed above. In contrast, if we fix
$m_{1/2}\simeq 1$~TeV, and take $m_0$ to be multi-TeV, MHDM is obtained
for values $|M_3|/m_{1/2} > 1$. To simplify fine tuning issues, we will
confine ourselves to $m_0 \lesssim 1$~TeV where we can obtain agreement with
(\ref{wmap1}) by reducing the value of $|M_3|$. We may expect an increase
in the $b$-multiplicity from SUSY events at the LHC because of the
enhanced higgsino content of the LSP.

\subsection{High ${\bf M_2}$ Dark Matter Model}

Very recently, it has been pointed out \cite{hm2dm} that raising the GUT
scale value of $M_2$ from its unified value of $m_{1/2}$ to about
(2.5--3)$m_{1/2}$ for $M_2 > 0$, or to between $-2$ and $-2.5$ times
$m_{1/2}$ for $M_2< 0$, also leads to a small value of $|\mu|$, giving
rise to a relic density in agreement with (\ref{wmap1}). The parameter
space of this high $|M_2|$ dark matter (HM2DM) model is given by,
\be
m_0, m_{1/2}, M_2, A_0, \tan\beta, {\rm sign}(\mu) \ \ ({\rm HM2DM})\;.
\ee
where $|M_2|$, the GUT scale value of the wino mass parameter, is
dialled to large magnitudes to obtain MHDM. The large value of $|M_2|$
causes the Higgs SSB $m_{H_u}^2$ to initially increase from its GUT
scale value of $m_0^2$ as $Q$ is reduced from $M_{\rm GUT}$. Ultimately,
however, the usual top quark Yukawa coupling effects take over, causing
$m_{H_u}^2$ to evolve to negative values resulting in the well-known
radiative breaking of electroweak symmetry. However, because of its
initial upward evolution, the weak scale value of $m_{H_u}^2$ is not as
negative as in models with unified gaugino masses, and the value of
$\mu^2$ is correspondingly smaller. The neutralino LSP then has a
significant higgsino component, and we may expect an enhancement of $b$-jets
in SUSY events at the LHC.

\section{Event simulation and calculational details} \label{sim}

We use ISAJET 7.74 \cite{isajet} with the toy calorimeter described in
Ref.~\cite{bcpt} for the calculation of the SUSY signal as well as of SM
backgrounds in the experimental environment of the LHC. We define jets
using a cone algorithm with a cone size $\Delta R
=\sqrt{\Delta\eta^2+\Delta\phi^2}= 0.7$. Hadronic clusters with $E_T >
40$~GeV and $|\eta({\rm jet})| <$ 3 are classifed as jets. Muons
(electrons) are classified as isolated if they have $E_T > 10$~GeV
(20~GeV) and visible activity in a cone with $\Delta R=0.3$ about the
lepton direction smaller than $E_T < 5$~GeV. We identify a hadronic
cluster with $E_T \ge 40$~GeV and $|\eta(j)|< 1.5$ as a $b$ jet if it
also has a $B$ hadron, with $p_T(B) > 15$~GeV and $|\eta(B)| < 3$,
within a cone with $\Delta R = 0.5$ of the jet axis. We conservatively
take the tagging efficiency\footnote{Notice that we assume that 50\% of
$b$-jets with $E_T>40$~GeV and in the central region will be
tagged. This is in contrast to a recent study \cite{india} where the
50\% efficiency refers to {\it all} $b$-jets. Effectively, the
efficiency in their study is significantly higher than in this
paper. Assuming that this larger tagging efficiency will be attained at
the LHC, these authors conclude that requiring 3 tagged $b$-jets
provides the best discrimination between the SM background and the SUSY
signal in the HB/FP region of the mSUGRA model.}  $\epsilon_b=0.5$ at the
LHC design luminosity of 100~fb$^{-1}$/y, and assume that gluon and
light quark jets can be rejected as $b$ jets by a factor $R_b= 150$ (50)
if $E_T < 100$~GeV ($E_T > 250$~GeV) and a linear interpolation in
between \cite{brej}. For jets not tagged as a $b$-jet, we require
$E_T(j) \ge 50$~GeV.

Gluino and squark production is the dominant sparticle production
mechanism at the LHC for gluino and squark masses up to about 1.8~TeV, if
$m_{\tq} \simeq m_{\tg}$. If instead squarks are very heavy, gluino pair
production will dominate the sparticle production rate up to about
$m_{\tg}\sim 0.8$~TeV. Cascade decays of the parent gluinos and squarks
then lead to signals in various multi-jet plus multi-lepton plus $\eslt$
topologies \cite{cascade}.  In some scenarios isolated
photons from radiative decays of neutralinos to lighter neutralinos or
to an ultra-light gravitino may also be present. Our focus, however, is
not on these scenarios, but instead on models of the type discussed in
Sec.~\ref{models} where $b$ and/or $t$ quarks are produced in these
cascades at a large rate.

Since SUSY particles are expected to be heavy (relative to SM particles)
sparticle production is expected to be signalled by events with hard
jets, possibly with hard, isolated leptons and large $\eslt$.  The
dominant physics backgrounds to these events with hard jets come from
$t\bar{t}$ production, $V + j$ production ($V=W, Z$), $VV$ production
and QCD production of light jets, where the $\eslt$ comes from neutrinos
produced by the decays of $W$ or $Z$ bosons or of heavy
flavours. Missing $E_T$ may also arise from mismeasurement of jet or
lepton transverse momenta and from uninstrumented regions of the detector.
These non-physics sources of $\eslt$ are detector-dependent, and only
qualitatively accounted for in our simulation with the toy
calorimeter. With the hard cuts that we use to obtain the reach, we
expect that the physics backgrounds will dominate the
difficult-to-simulate detector-dependent backgrounds, and the results of
our analyses of the SUSY reach will be reliable. This expectation
is indeed borne out since results of previous theoretical analyses of
the SUSY reach \cite{bcpt,update} compare well with the projected
reaches obtained by the CMS \cite{cms} and ATLAS \cite{atlas}
collaborations. The {\it gain in reach}, if any, that we obtain from
$b$-jet tagging, should if anything be more reliable than the absolute
value of the reach.\footnote{The absolute reach may also suffer from the
fact that SM backgrounds may be somewhat larger than those obtained
using shower Monte-Carlo programs when proper matrix elements for multi-jet
production are included. We expect though that the gain in the reach
from $b$-tagging may again be less sensitive to the inclusion of the
proper matrix elements.}

In the analysis detailed in the next section, we have examined the reach
of the LHC for a wide range of sparticle masses, for the different
models introduced in Sec.~\ref{models}. To facilitate this, we generate
signals and backgrounds (calculational details are described below) and
only write out events that include at least two jets with $E_T(j) \ge
100$~GeV and $\eslt \ge 100$~GeV, which we refer to as our basic cuts.
\begin{table}[htdp]
\begin{center}
\begin{tabular}{lcccc}
\hline\hline
Source & $\sigma_{\rm basic}$  & $\sigma_{\rm cut}(0b)$  
&$\sigma_{\rm cut}(1b)$  & $\sigma_{\rm cut}(2b)$\\ \hline
$t\bar{t}$ &19900 &2.16 & 1.41 & 0.365 \\
$W+j$ &21400 & 12.0 & 1.36 & 0.133 \\
$Z+j$ & 8850 & 5.11 & 0.059 & 0.0052 \\
$VV$  &89.8 & 0.0248 & 0.0020 & 0.0001 \\
QCD & 93700 &  11.6 & 3.11 & 0.467\\
Total &$1.44\times 10^5$ & 30.9 & 5.94 & 0.97 \\
mSUGRA1 & 261 & 12.0 &9.26 & 3.86\\
mSUGRA2 & 48.4 & 2.44 &1.95 & 0.87\\
\hline\hline
\end{tabular}
\end{center}
\caption{Cross sections in fb for the SM production of $t\bar{t}$,
  $W+j$, $Z+j$, $VV$, and QCD jet events that form the dominant
  backgrounds to the multi-jet plus $\eslt$ signal from sparticle
  production at the LHC. The second column gives the cross section for
  events with the basic requirements of two jets with $E_T(j) \ge
  100$~GeV and $\eslt \ge 100$~GeV. The last three columns give the
  corresponding cross sections for the softest of the final set of cuts
  (listed in the bottom part of Table~\ref{tab:cuts}) that we actually use
  in our analysis, with no requirement of $b$-jet tagging (column 3),
  requiring at least one tagged $b$-jet (column 4) and at least two
  tagged $b$-jets (column 5). For illustration, we also list the
  corresponding signal cross sections for two points in the HB/FP region
  of the mSUGRA model, with $A_0=0$, $\tan\beta=10$ and $m_{\tg}\simeq
  1$~TeV, $m_{\tq}\sim 3$~TeV (mSUGRA1) and $m_{\tg}\simeq 1.5$~TeV and
  $m_{\tq}\sim 3.9$~TeV (mSUGRA2). }
\label{tab:bkg}
\end{table}
The corresponding cross sections for SM events are shown in the second
column of Table~\ref{tab:bkg}.  For low to medium values of sparticle
masses, the sparticle production cross sections are large enough for us
to extract the signal above SM backgrounds with relatively soft analysis
cuts. For very heavy sparticles, however, the production rate is
small, but essentially all events contain very energetic jets and large
$\eslt$. The detection of the signal is then optimized by using very
hard cuts that strongly suppress SM backgrounds while retaining bulk of
the SUSY signal.  
Since our aim is to develop a strategy that can be applied to
essentially the entire interesting mass range of a wide variety of
models, 
we are led to evaluate the signal together with
the SM background for a wide range of cuts, detailed in the next
section. To understand the relative importance of the different
background sources, in the last three columns of Table~\ref{tab:bkg} we
list the corresponding cross sections for the {\it softest} set of cuts
that we use in our analysis detailed in Sec.~\ref{sec:btag}.

In the last two rows we also list the corresponding signal cross
sections for two WMAP-consistent cases in the HB/FP region of the mSUGRA
model. Several comments are worth noting.
\begin{itemize}
\item We see that with the basic requirements of two jets with $E_T \ge
  100$~GeV and $\eslt \ge 100$~GeV, the background is two (three) orders
  of magnitude larger than the signal  for
  $m_{\tg} \simeq 1$ (1.5)~TeV; however, the analysis cuts very
  efficiently reduce the background, while reducing the signal by a much
  smaller factor. 

\item After these analysis cuts we see that QCD, followed by $V+j$
production, are the leading backgrounds to the inclusive $\eslt$
signal. Top pair production, while significant, is considerably
smaller. Since we do not require the presence of leptons, the background
from $VV$ production is negligible.

\item The backgrounds from QCD and $V+j$ production can be sharply
  reduced by the use of $b$-jet tagging with relatively small loss of
  the signal. In contrast, since top events necessarily contain
  $b$-jets, $b$-tagging reduces the $t\bar{t}$ background only by a
  modest amount.
\end{itemize}

Table~\ref{tab:bkg} highlights the importance of a careful evaluation of
the QCD and the $V+j$ backgrounds. This is technically complicated
because the large size of the cross sections necessitates simulations of
very large number of events to obtain a reliable estimate for the
backgrounds after the very hard cuts that are needed for optimizing the
reach of the LHC.\footnote{Of course, the fact that we are far into the
tails of these backgrounds where the simulations (which will be tuned to
data when these become available) require possibly unjustified
extrapolations is a different matter.} Moreover, since the cross section
is a rapidly falling function of the centre of mass energy, or
equivalently, the hard-scattering $p_T$ of the initial partons, we must
ensure that our procedure generates events even for very large values of
$P_T^{HS}$ where the matrix element is very small, so that these events
which have  much smaller weights are included in the analysis. To
facilitate this, we have generated the various backgrounds using
different numbers, $N_i^{HS}$, of hard scattering bins: the bin
intervals are finely spaced for low values of $P_T^{HS}$ where event
weights are very large. We choose $N_i^{HS}=53, 13, 8$ and 7 for $i =$
QCD, $V+j$, $t\bar{t}$ and $VV$, respectively, where the choice
$N_{HS}^{\rm QCD} = 53$ reflects the largeness of the QCD cross
section. We have generated a total of about 10M QCD events, about 1M
$W+j$ events and about 500K-700K events for each of the other
backgrounds.  If, for any set of cuts, we find zero events in our
simulation of a particular background, we set this background cross
section to a value corresponding to the one event level in the bin with
the smallest weight in our simulation.

\section{Bottom jet tagging and the reach of the LHC} \label{sec:btag}

\subsection{Simulation of the Signal and the LHC reach} 
Simulation of the signal events is technically much easier than that of
the background. This is largely because the signal typically originates 
in heavy sparticles, and so passes the hard analysis cuts with relative
ease compared to the background. To assess how much $b$-jet tagging
extends the SUSY reach of any particular model, rather than perform
extensive and time-consuming scans of the parameter space, we have
defined ``model lines'' along which the sparticle mass scale increases. 
We then choose parameters along these lines, and for every such
parameter set use ISAJET 7.74 to generate a SUSY event sample. Next, we
pass this event sample through the set of analysis cuts defined
below, and define the signal to be observable at the LHC if for {\it any}
choice of cuts
\bi
\item the signal exceeds 10 events, assuming an integrated luminosity of
  100~fb$^{-1}$, 
\item the statistical significance of the signal $N_{\rm
  signal}/\sqrt{N_{\rm back}}\ge 5$,
  and
\item the signal to background ratio, $N_{\rm
  signal}/N_{\rm back} \ge 0.25$.  
\ei 
We also require a minimum of 15 events after cuts in our simulation of
the signal. We obtain the reach for each model line by comparing the
corresponding signal with the background, and ascertaining where the
signal just fails our observability criteria for {\it the entire set of
cuts in Table~\ref{tab:cuts}} below. 

\subsection{Analysis cuts} \label{sec:cuts}

The inverted mass hierarchy model based on $SO(10)$ SUSY GUTs, whose hallmark
is the light third generation, serves as the prototypical case  where we
expect enhanced $b$-jet multiplicity in SUSY events. We have used this
framework to guide us to the set of analysis cuts that can be used for
the optimization of the SUSY signal for a wide range of sparticle masses
in a wide class of models. Toward this end, we fix $\mu < 0$, $A_0<0$,
and $\tan\beta=47$ (a large value is needed for the unification of
Yukawa couplings) and choose $m_{10} =\sqrt{2}m_{16}$, $A_0=-2m_{16}$ to
obtain the hierarchy between the first/second and third generation
scalars as discussed above. The choice $M_D = 0.25m_{16}$ facilitates
electroweak symmetry breaking. We vary the gluino mass along the ``model
line'' with $m_{1/2}=0.36m_0+48$~GeV which maintains a hierarchy between
the generations. The value of
$$ S \equiv \frac{3(m_{\tu_L}^2 + m_{\td_L}^2 + m_{\tu_R}^2 +
m_{\td_R}^2) + m_{\te_L}^2 + m_{\te_R}^2 + m_{\tnu_e}^2}{3(m_{\tst_1}^2
+ m_{\tb_1}^2 + m_{\tst_2}^2 + m_{\tb_2}^2) + m_{\ttau_1}^2 +
m_{\ttau_2}^2 + m_{\tnu_\tau}^2} 
$$
is typically around 3.5-4.1 along this model line. 

The optimal choice of cuts depends on the ({\it a priori} unknown)
sparticle spectrum, and to a smaller extent on their decay patterns. While
hard cuts optimize the signal if sparticles are heavy, these would
drastically reduce (or even eliminate) the signal if sparticles happen
to be light. In order to obtain a general strategy that can be used for
a wide variety of models, we have used the $SO(10)$ model with $\mu < 0$
to devise a universal set of cuts that can be used for SUSY discovery in
any of the various models that we have introduced, and likely, also for
a wider class of models.

Toward this end, we generate a sample of signal events for this ``test model
line'' and run this, as well as the SM backgrounds that we discussed
above, through each one of the large set of analysis cuts
detailed in the upper part of
Table~\ref{tab:cuts}. Here, $m_{\rm eff}$ is the scalar sum of the
transverse energies of the four hardest jets in the event combined with
the missing transverse energy, $\Delta\phi$ is the transverse plane
opening angle between the two hardest jets, and $\Delta\phi_b$ the
corresponding angle between the two tagged $b$-jets in events with
$n_b\ge 2$. To clarify, the softest set of cuts that we use for the $0b$
signal has [$\eslt$, $E_T(j_1)$, $E_T(j_2)$, $E_T(b_1)$, $m_{\rm eff}$]
$\ge [300, 300, 100, 40, 1500]$~GeV, $n_j \ge 4$ and transverse
sphericity $S_T >0.1$, with no
restriction on jet opening angles.
Next, we harden the cut on one of these observables to the next level,
keeping the others at the same value, {\it etc.} until the complete set
of $6\times 5\times 6\times 3\times 3\times 5\times 2 \times 21$
combinations has been examined for $n_b\ge 2$. Since there are (is) no
(just one) tagged $b$ jets in the $n_b=0$ ($n_b=1$) case, there are
correspondingly fewer combinations for these analyses. 
\begin{table*}[htdp]
\begin{center}
\begin{tabular}{lcc} \hline \hline
 Variable  & $0b$, $1b$ & $2b$ \\  \hline
$\eslt$~(GeV) $>$ & $300, 450, 600, 750, 900, 1050$ &
$300, 450, 600, 750, 900, 1050$ \\ 

$E_T(b_1)$~(GeV) $>$ & $40, 100, 200, 300, 400$ & $40, 100, 200, 300, 400$\\
$m_{\rm eff}$~(GeV) $>$ & $1500, 2000, 2500,..., 4000$ &
$1500, 1750, 2000, 2250, 2500, 2750$\\ 
$\Delta\phi < $ & $180^\circ$, $160^\circ$,
$140^\circ$ & $180^\circ$, $160^\circ$, $140^\circ$ \\
$\Delta\phi_{b}< $ & n/a & $180^\circ$, $150^\circ$, $120^\circ$\\
$n_j\ge$ & $4, 5, 6, 7, 8$ & $4, 5, 6, 7, 8$ \\ 
$S_T\ge$ & $0.1, 0.2$ & $0.1, 0.2$\\ \hline 
$[E_T(j_1), E_T(j_2)]$~(GeV) $>$ &
\multicolumn{2}{c} 
{$(300,100), (300, 200), (400, 200), (400,300),
(500, 200), (500, 300), $} \\
& \multicolumn{2}{c}{$ (500, 400), (600, 200), 
(600,300), (600,400), (600,500), $
}\\
& \multicolumn{2}{c}{$ (700,200), (700,300), 
(700,400), (700,500), (700,600), $
}\\
& \multicolumn{2}{c}{$ (800,200), (800,300), 
(800,400), (800,500), (800,600) $
}\\
\hline \hline
\multicolumn{3}{c} {\bf{Final Cuts}} \\ \hline\hline
$\eslt$~(GeV) $>$ & $450, 600, 750, 900, $ &
$450, 600, 750$ \\ 
 
$E_T(b_1)$~(GeV) $>$ & $40, 100, 200$ & $40, 100, 200, 300$\\
$m_{\rm eff}$~(GeV) $>$ & $1500, 2000, 2500,..., 4000$ &
$1500, 1750, 2000, 2250$\\ 
$\Delta\phi < $ & $180^\circ$, $160^\circ$,
$140^\circ$ & $180^\circ$ \\
$\Delta\phi_{b}< $ & n/a & $180^\circ$, $150^\circ$, $120^\circ$\\
$n_j\ge$ & $4, 5, 6, 7, 8$ & $4, 5, 6, 7$ \\ 
$S_T\ge$ & $0.1$ & $0.1$\\ \hline 
$[E_T(j_1), E_T(j_2)]$~(GeV) $>$ &
\multicolumn{2}{c} 
{$ (300, 200), (400, 200),
(500, 200), (500, 300), (500, 400), (600, 200),$} \\
& \multicolumn{2}{c}{$   
(600,500), (700,300), (700,600), (800,300), (800,600) $
}\\
\hline \hline
\end{tabular}
\end{center}
\caption{The complete set of cuts that we examined for
extraction of the SUSY signal over the SM backgrounds is shown in 
the upper part of the Table. 
 The $0b$, $1b$ and $2b$
entries respectively denote requirements for events without any
restriction $b$-jet tagging, with at least one tagged $b$-jet, and with
at least two tagged $b$-jets.
The lower part of the Table shows the final set of
cuts that we recommend for the extraction of the SUSY signal over the
entire range of massses and models that we have explored in the paper. 
}
\label{tab:cuts}
\end{table*}

For each of these cut choices, we analysed the observability and
statistical significance of the LHC signal for our test $SO(10)$ model
line for an integrated luminosity of 100~fb$^{-1}$. We found that the
subset of cuts shown in the lower part of Table~\ref{tab:cuts} (and
labelled ``Final Cuts'') 
is sufficient to ensure the observability of the
SUSY signal over the entire mass range. Restricting the analysis
to this subset  has no impact on either the observability or the
statistical significance of the signal over the entire sparticle mass
range. In the remainder of this paper we, therefore, confine ourselves
to this limited subset of cuts, as this speeds up the analysis considerably.

\subsection{Results}
In this section, we evaluate prospects for increasing the reach of the
LHC by the use of $b$-tagging to reduce SM backgrounds, thereby
increasing the statistical significance of the SUSY signal, for each of
the models introduced in Sec.~\ref{models}. We confine ourselves to
various 1-parameter model lines (introduced below) along which sparticle
masses increase and run the signal and backgrounds through each of the
final set of cuts in Table~\ref{tab:cuts}, and optimize the signal by
selecting the cut choice that yields an observable signal with the
highest statistical significance.  To assess the gain from $b$-tagging,
for each model line we first do so without any requirement on
$b$-tagging, and then repeat it requiring, in addition, at least one and at least two tagged $b$-jets.

\subsubsection{The HB/FP region of the mSUGRA model} 

The possibility of increasing the LHC reach was first studied in the
HB/FP region of the mSUGRA framework \cite{MMT}, where it was found
that the reach could be increased by up to 15-20\%. We have repeated
this study, albeit with a somewhat different model line with 
$$m_{1/2}= 0.295 m_0 - 507.5~{\rm GeV}, \tan\beta=30, A_0=0,$$
in the HB/FP
region that saturates the relic density in (\ref{wmap1}) and of course,
with the different set of cuts that we use here. We  find an
increased reach from $b$-tagging in qualitative agreement with
Ref.~\cite{MMT}. 

\subsubsection{Inverted mass hierarchy model} \label{imhmodel}

As discussed in Sec.~\ref{sec:cuts}, we have already used the $SO(10)$
model with $\mu <0$ and parameters related by (\ref{so10rel}) where we
obtain an  inverted mass hierarchy to choose the final set of cuts for our
analysis. Here, we show results for the reach of the LHC with and
without requirements of $b$-jet tagging for two model lines with a
significant inversion of the sfermion mass hierarchy, one for
each sign of $\mu$. For both of these, we choose
\begin{equation}
-A_0=2m_{16}=\sqrt{2}m_{10}, \tan\beta=47, \label{above} \end{equation} 
with
\begin{eqnarray}
M_{D} = 0.25m_{16} \ {\rm and} \ m_{1/2} = 0.36m_{16} + 48~{\rm GeV}
\ {\rm for} \ 
\mu <0,\nonumber\\  \label{imhneg}  \\
M_{D} = 0.20m_{16}  \ {\rm and} \ m_{1/2} = 0.30m_{16} + 39~{\rm GeV} \
{\rm for} \
\mu >0.\nonumber \\  \label{imhpos}
\end{eqnarray}
Of course, to determine the reach for the $\mu <0$ model line we
generate different sets of signal events from those used in
Sec.~\ref{sec:cuts}. 

\begin{figure*}[htb]{\begin{center}\vspace{-1cm}
\epsfig{file=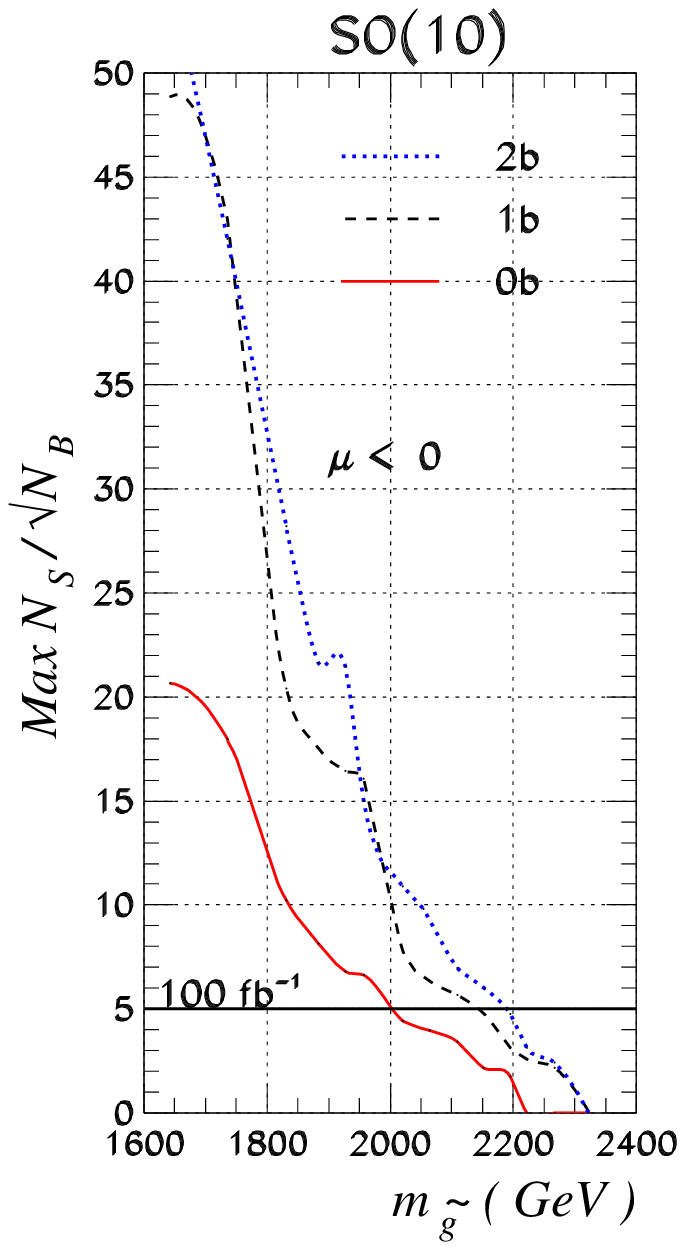,width=7.6cm}\hspace{-3cm} 
\epsfig{file=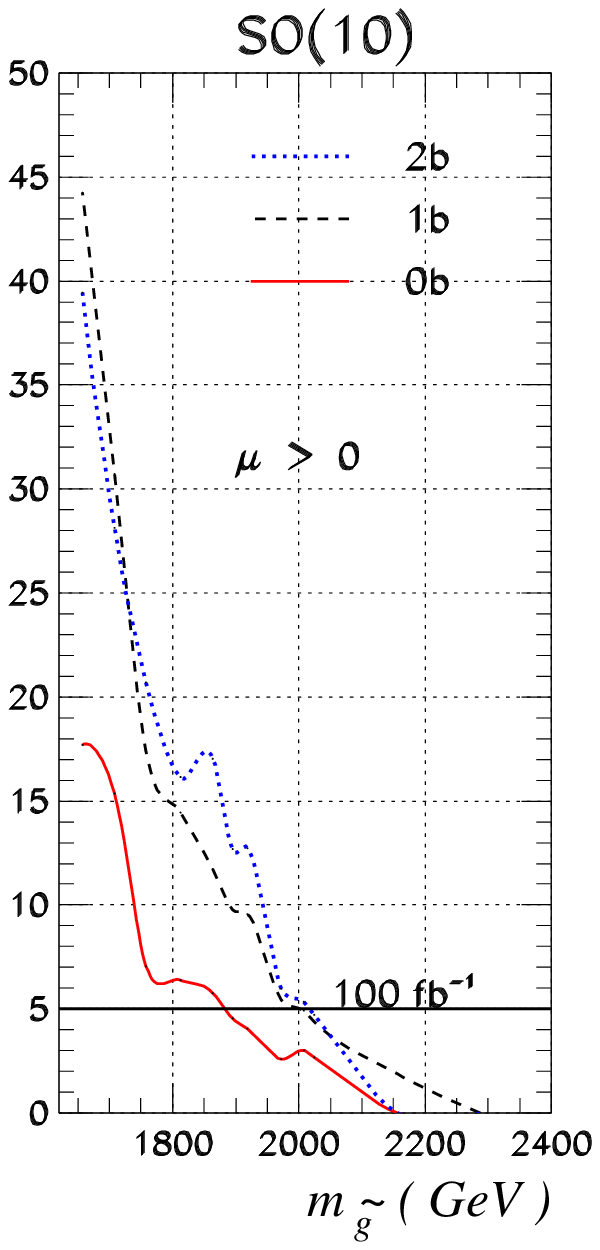,width=7.6cm} 
\caption{\label{reach:so10} The statistical
significance  of the SUSY signal
satisfying our observability criteria at the LHC for the
  inverted hierarchy $SO(10)$ model lines introduced in the text, assuming an
  integrated luminosity of 100~fb$^{-1}$ for  ({\it a})~$\mu <0$, and
  ({\it b})~$\mu>0$. The solid (red) line is for the signal with no
  requirement on $b$-tagging, the dashed (black) line is with the
  requirement of at least one tagged $b$-jet, and the dotted (blue) line
  is with at least two tagged $b$-jets. The signal is observable if the
  statistical significance is above the horizontal line at $N_{\rm
  signal}/\sqrt{N_{\rm back}}=5$. }\end{center}}
\end{figure*}

Our results are shown in Fig.~\ref{reach:so10}, where we plot the largest
statistical significance of the signal, $N_{\rm signal}/\sqrt{N_{\rm back}}$,
versus the
corresponding gluino mass for ({\it a})~$\mu <0$, and ({\it b})~$\mu
>0$, assuming an integrated luminosity of 100~fb$^{-1}$. The maximal
$N_{\rm signal}/\sqrt{N_{\rm back}}$ was obtained running over all the
cuts in Table~\ref{tab:cuts},  subject to the requirement that the
$N_{\rm signal}/N_{\rm back}>0.25$ and $N_{\rm signal}>10$ 
event criteria are satisfied. The solid (red)
curves show this significance for the inclusive $\eslt$ signal with no
requirement of $b$-jet tagging, while the dashed (black) curve and the
dotted (blue) curves correspond to cases where we require at least one
and two tagged $b$-jets, respectively. The wiggles in these curves
reflect the statistical errors in our simulation. 
We attribute the 
somewhat larger reach in the left frame to the fact that 
the mass hierarchy (as measured by the value of $S$) is somewhat smaller
for $\mu<0$, so that $\tq\tg$ makes a larger contribution in this case. 
We also see that for $\mu
<0$, $b$-tagging leads to an increase of the LHC reach by $\sim
200$~GeV, or about 10\%, while the corresponding increase is somewhat
smaller for the model line with positive $\mu$. 
%We have traced this
%difference to the fact that a larger hierarchy (as measured by the
%value of $S$) is obtained for $\mu<0$ than for positive values of
%$\mu$. Since the presence of $b$-jets in SUSY events is a direct result
%of the relative smallness of third generation squark masses, it is not
%surprising that the LHC reach is more enhanced for $\mu<0$.
This difference (which may well not be very significant in view of the
wiggles) is evidently due to the increased reach in the $2b$ channel,
and 
could arise from a complicated interplay between the effect of cuts and
the sparticle spectrum: for instance, for $m_{\tg}\sim 1960$~GeV,
$m_{\tb_1}$ is significantly lighter in the $\mu<0$ case, while
$m_{\tst_1}$ is considerably heavier. As a result, the branching
fraction for the decays $\tg \to b\tb_i$,
which likely leads to a harder spectrum for $b$-jets (compared to
$\tg \to t\tst_1$, which constitutes the bulk of the remaining
decays of the gluino), falls from 38\% for
negative $\mu$ to 28\% for positive $\mu$.

\subsubsection{Non-universal Higgs mass models} 

Next, we turn to the impact of $b$-tagging on the reach in NUHM models
with just one additional parameter $m_\phi$ that is adjusted so that
agreement with the observed relic density is obtained either by tempering
the LSP content so that it is MHDM ($m_\phi>m_0$), or by adjusting the masses
so that the LSP annihilation rate is resonantly enhanced by the exchange
of neutral $A$ or $H$ bosons in the $s$-channel ($m_\phi<0$). We did not
study the NUHM model where both Higgs SSB mass parameters are
arbitrary -- the so-called NUHM2 models in the nomenclature of
Ref.~\cite{NUHM} -- because this meant that both $m_A$  and $\mu$ are arbitrary, resulting in too much freedom for definitive analysis. 
Beginning with the MHDM cases of the LSP where sparticle decays to third
generation quarks are enhanced by the higgsino content of the LSP, we
introduce two model lines with $A_0=0$, $\tan\beta=10$ and $\mu>0$, with
(1)~$m_0= m_{1/2}$, and (2)~$m_0=3m_{1/2}$, for which we have to choose
$m_\phi\simeq 1.7m_0$ and $m_\phi\simeq 1.12m_0$, respectively, in order
to obtain the observed relic density. In the former case, the
squarks of the first two generations are roughly degenerate with
gluinos, whereas in the latter case $m_{\tq}\sim 1.6 m_{\tg}$. 
 \begin{figure*}[htb]{\vspace{-1cm}\begin{tabular}{lll}
\epsfig{file=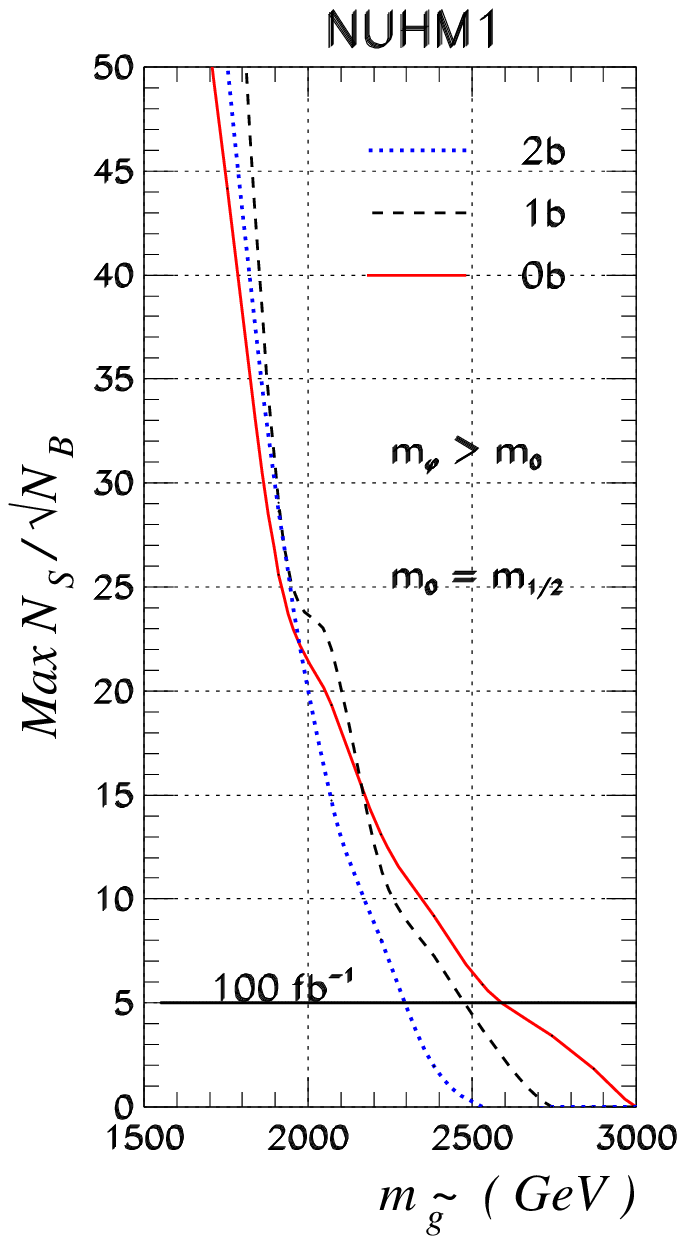,width=7.6cm}\hspace{-2.9cm}  
\epsfig{file=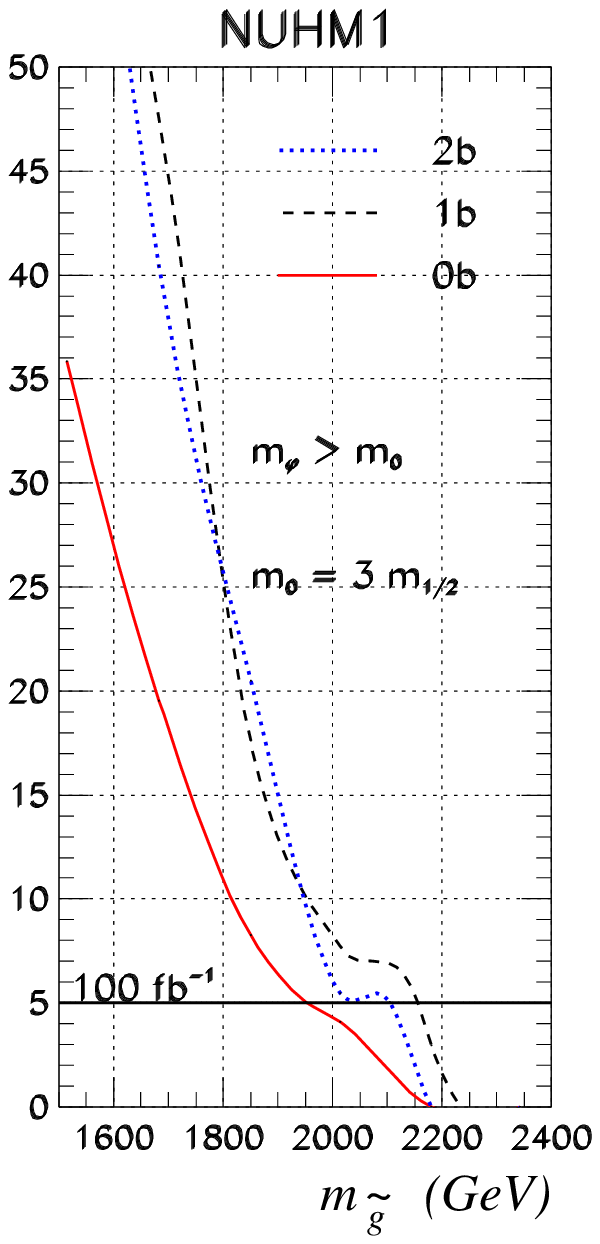,width=7.6cm}\hspace{-2.9cm} 
\epsfig{file=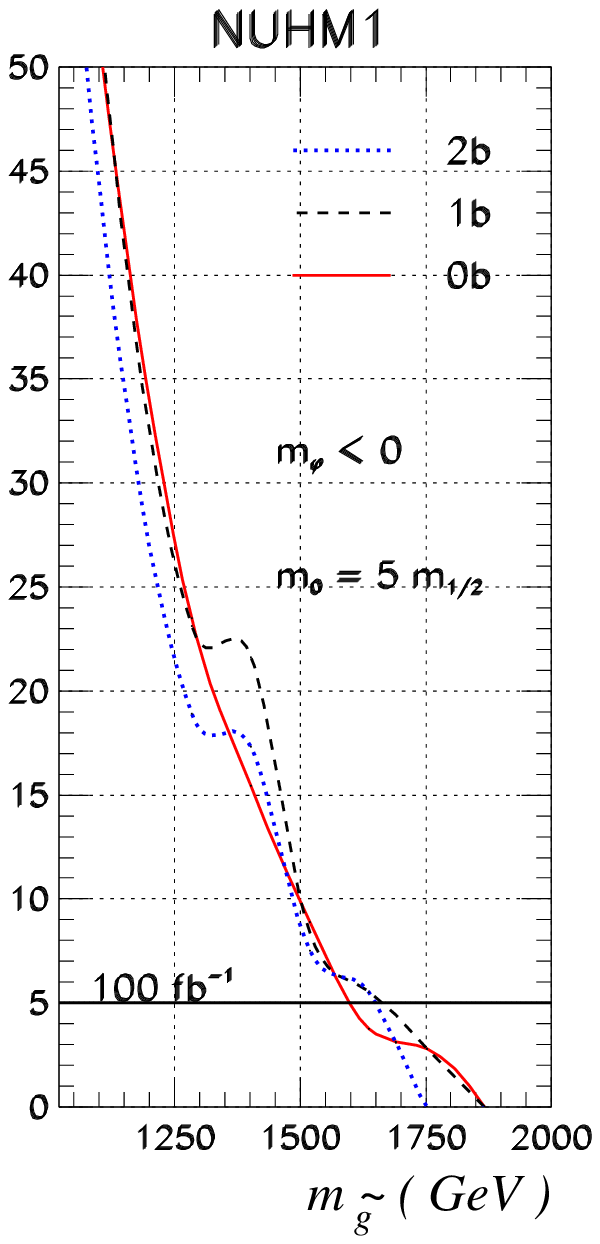,width=7.6cm}
\end{tabular}
\caption{\label{reach:nuhm} The statistical significance of the SUSY
signal satisfying our observability criteria at the LHC for the three
NUHM model lines introduced in the text, assuming an integrated
luminosity of 100~fb$^{-1}$. All the model lines have $A_0=0$ and
$\mu>0$, with ({\it a})~$m_\phi >0, \tan\beta=10, m_0=m_{1/2}$, ({\it
b})~$m_\phi >0, \tan\beta=10, m_0=3m_{1/2}$, and ({\it c})~$m_\phi <0,
\tan\beta=20, m_0=5m_{1/2}$. The solid (red) line is for the signal with
no requirement on $b$-tagging, the dashed (black) line is with the
requirement of at least one tagged $b$-jet, and the dotted (blue) line
is with at least two tagged $b$-jets. The signal is observable if the
statistical significance is above the horizontal line at $N_{\rm
signal}/\sqrt{N_{\rm back}}=5$. }}
\end{figure*}

Our results for the statistical significance of the LHC SUSY signal,
with and without $b$-jet tagging are shown in Fig.~\ref{reach:nuhm} for
({\it a})~$m_0=m_{1/2}$, and ({\it b})~$m_0=3m_{1/2}$. We see that while
$b$-tagging clearly improves the reach by $\sim 10\%$ in the case shown
in frame ({\it b}), it leads to a {\it degradation} of the reach in
frame ({\it a}).  We have traced this to the fact that for this case
where squark and gluino masses are comparable, squark production
(particularly first generation squark production) makes a significant
contribution to the signal after the hard cuts. Then, since these
squarks dominantly decay to
charginos and neutralinos (remember that because $m_{\tq}\sim m_{\tg}$,
the decay $\tq \to q\tg$ is suppressed by phase-space) plus 
{\it quarks of their own generation} there are essentially no $b$-quarks
produced in squark decays, and a sizeable fraction of the
inclusive $\eslt$ signal is actually cut out by any $b$-tagging
requirement. In frame ({\it b}), the squarks are much heavier than
gluinos and so contribute a smaller fraction of the signal, but more
relevantly, $\tq \to q\tg$ with a large branching fraction, so
that $b$-tagging helps in this case. These considerations also explain
why the increase in reach from $b$-tagging is not as large as in the
case of the HB/FP region of the mSUGRA model where $m_{\tq} \gg m_{\tg}$
\cite{MMT}.

We now turn to the $m_\phi <0$ model line shown in
Fig.~\ref{reach:nuhm}{\it c} for which we have chosen $m_0=5m_{1/2}$ (to
ensure squark contributions to the signal do not dilute the effect of
$b$ tagging as in the case that we just discussed), $A_0=0$,
$\tan\beta=20$ and $\mu>0$, and $m_{\phi}$ is adjusted to be 
about $-1.47m_0$ to give
agreement with (\ref{wmap1}) via resonant annihilation of LSPs through
$A/H$ exchanges in the $s$-channel. This means that $A$ and $H$ must be
relatively light and accessible in cascade decays of gluinos and
squarks.  However, we see no enhancement of the LHC reach in this
case. We understand this in hindsight. In this case $|\mu|$ is large so
the lighter neutralinos produced in gluino cascade decays are
gaugino-like, with $m_{\tw_1}\simeq m_{\tz_2}\simeq 2m_{\tz_1}$. Then
the very condition $2m_{\tz_1}\sim m_A$ that makes the LSP anihilation
cross section resonant suppresses the phase space for the decays of
$\tz_2 \to A \ {\rm or} \ H + \tz_1$, so that these are not
significantly produced in cascade decays of gluinos. Since squarks are
very heavy, they are essentially irrelevant to this discussion.

\subsubsection{Low $M_3$ dark matter model} 

As explained above, we can also obtain MHDM, and hence a potential
increase in reach via $b$-tagging, in models with non-universal gaugino
%
%\FIGURE[htb]{
\begin{figure}[htb]{\vspace{-0.5cm}\begin{center}
\epsfig{file=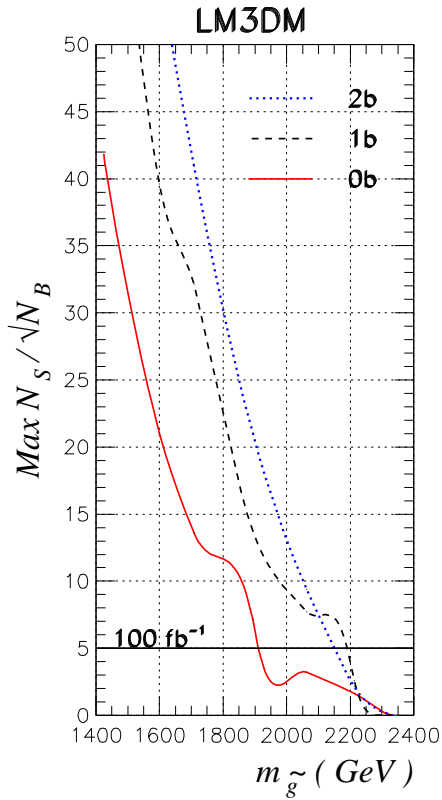,width=7.6cm} \end{center}
\caption{\label{reach:lm3dm} The statistical significance of the SUSY
 signal satisfying our observability criteria at the LHC for the LM3DM
 model line with $m_0=m_{1/2}, A_0=0, \tan\beta=10$ and $\mu>0$, where
 $M_3({\rm GUT})$ is adjusted to saturate the measured CDM relic
 density, assuming an integrated luminosity of 100~fb$^{-1}$.  The solid
 (red) line is for the signal with no requirement on $b$-tagging, the
 dashed (black) line is with the requirement of at least one tagged
 $b$-jet, and the dotted (blue) line is with at least two tagged
 $b$-jets. The signal is observable if the statistical significance is
 above the horizontal line at $N_{\rm signal}/\sqrt{N_{\rm back}}=5$. }}
\end{figure}
mass parameters where $|M_3({\rm GUT})|$ is taken to be reduced compared
to its value in models with gaugino mass unification. To study the gain
in the reach that we may obtain in this case, we have explored an LM3DM
model line with 
$$m_0=m_{1/2}, A_0=0, \tan\beta=10, \mu >0, $$
where the GUT scale value of $M_3$ (which we take to be positive) is
adjusted to saturate the measured CDM relic density.\footnote{Roughly
  speaking, for $m_0=m_{1/2}=700$~GeV, $M_3({\rm GUT})=277$~GeV, and for
  an increase of $\delta m_0$ in $m_0=m_{1/2}$, the GUT scale value of
  $M_3$ has to be raised by about $\delta M_3 \sim \delta m_0/2.25$. }

The corresponding dependence of the statistical significance of the SUSY
signal on $m_{\tg}$ is shown in Fig.~\ref{reach:lm3dm}. We see that in this
case $b$-tagging leads to an increase in reach close to 15\%. This is
because though gluinos and squarks are both reduced in mass relative to
their uncoloured cousins, the reduced value of the gluino mass parameter
leads to $m_{\tq}\sim (1.4-1.5)m_{\tg}$ even for $m_0=m_{1/2}$, to be
compared to $m_{\tq}\sim m_{\tg}$ that we obtained for models with unified 
gaugino masses as {\it e.g.} in  the NUHM case just discussed. The
large value of $m_{\tq}$ relative to $m_{\tg}$ then leads to an
enhanced reach via $b$-tagging just as before.

%
%\FIGURE[bth]{
\begin{figure*}[h]{\vspace{-1cm}\begin{center}\begin{tabular}{cc}
\epsfig{file=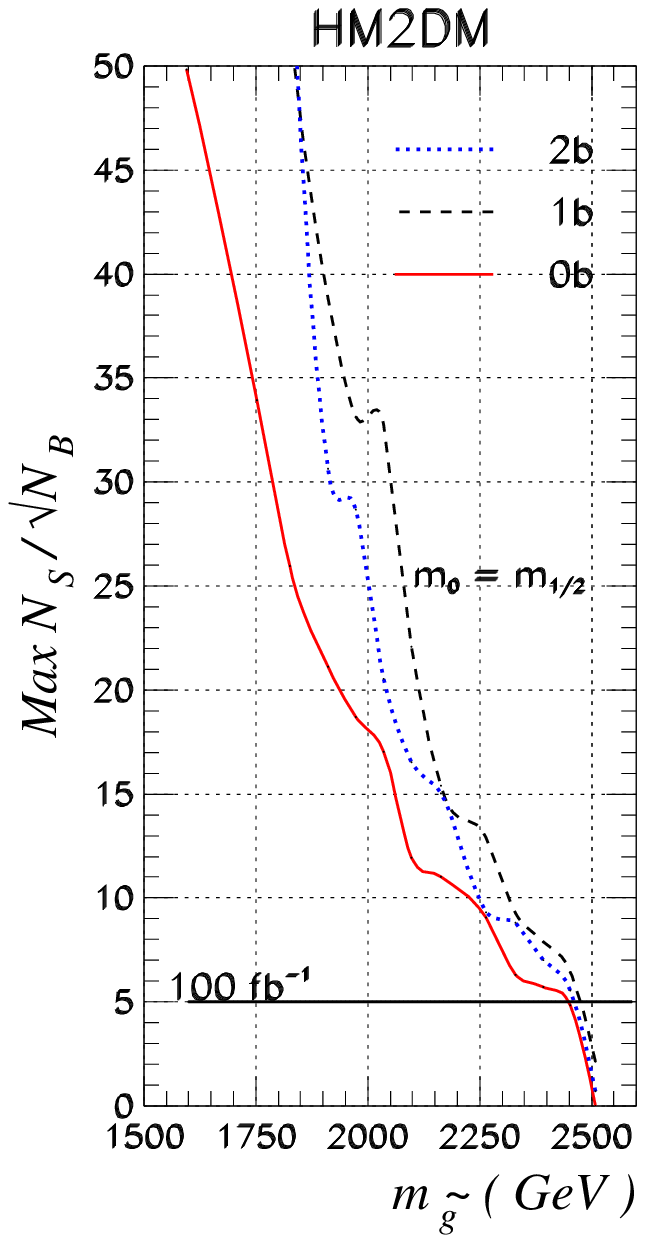,width=7.6cm}\hspace{-3cm}
 \epsfig{file=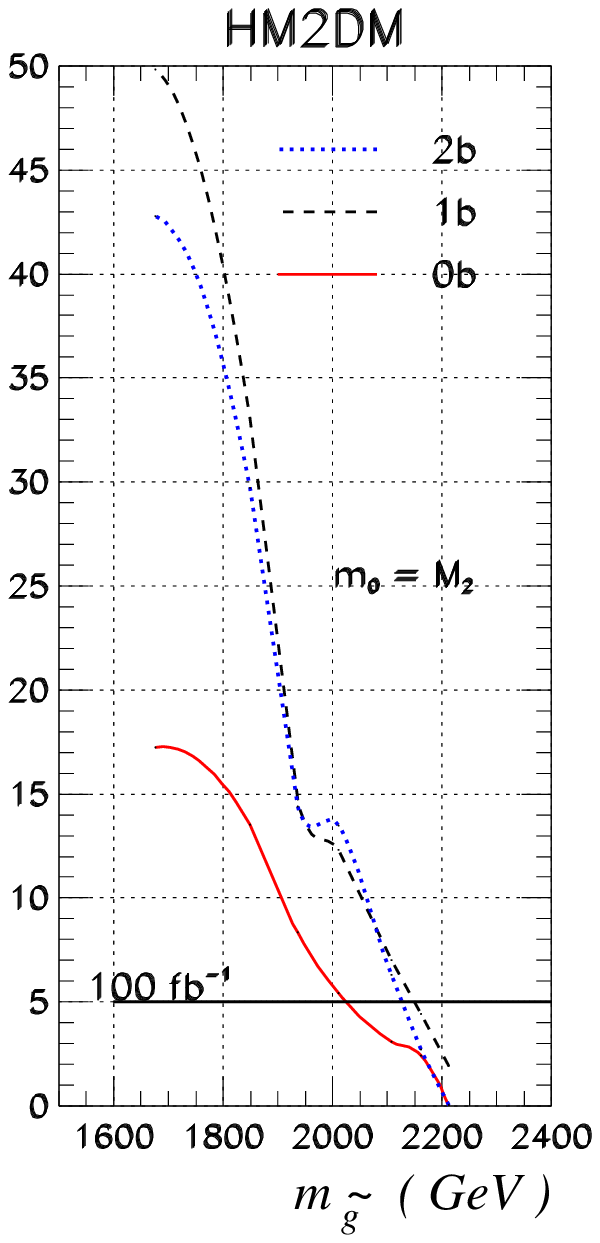,width=7.6cm} \end{tabular}\end{center}
\caption{\label{reach:hm2dm} The statistical significance 
 of the SUSY
signal satisfying our observability criteria at the LHC
for the
  HM2DM  model line with $A_0=0, \tan\beta=10$,
  $\mu>0$ and ({\it a})~$m_0=m_{1/2}$, and ({\it b})~$m_0=M_2({\rm
  GUT})$.
 In both frames, $M_2({\rm GUT})$ is adjusted to a positive value
  so as to saturate the measured
  CDM relic density, and an 
  an integrated luminosity of 100~fb$^{-1}$ is assumed.  The solid (red)
  line is for the signal with no requirement on $b$-tagging, the dashed
  (black) line is with the requirement of at least one tagged $b$-jet,
  and the dotted (blue) line is with at least two tagged $b$-jets. The
  signal is observable if the statistical significance is above the
  horizontal line at $N_{\rm signal}/\sqrt{N_{\rm back}}=5$. }}
\end{figure*}

\subsubsection{High $M_2$ dark matter model}

As a final example, we consider the LHC reach in the HM2DM model, where
agreement with (\ref{wmap1}) is obtained by raising $|M_2({\rm GUT})|$
from its canonical value of $m_{1/2}$ in models with gaugino mass
unification, so that the lightest neutralino is MHDM. Since the LSP
contains a substantial higgsino component, it is again reasonable to
expect that $b$-jet tagging may increase the SUSY reach of the LHC.  As
we have already seen in other examples, the increased reach from $b$-jet
tagging depends on the value of the squark mass relative to
$m_{\tg}$. This led us to consider two model lines with, ({\it
a})~$m_0=m_{1/2}$, and ({\it b})~$m_0=M_2({\rm GUT})$, for both of which
we take $\tan\beta=10$, $\mu >0$ and $A_0=0$. Since the correct relic
density is obtained by {\it raising} $M_2$, model-line ({\it b}) which
gives heavier squarks than model line ({\it a}) will give a smaller
reach as measured in terms of $m_{\tg}$. The {\it increase} in the reach
from $b$-jet tagging will, however, be larger for model line ({\it b})
since squark contributions to sparticle production are kinematically
suppressed. 

The statistical significance of the SUSY signal in the HM2DM model is
shown for the two model lines in the two frames of Fig.~\ref{reach:hm2dm}. 
Indeed we see that while the reach in the left frame for $m_0=m_{1/2}$
extends to $m_{\tg} \leq 2.5$~TeV (as compared to 2.1~TeV in the right
frame), there is very little gain in the reach from $b$-jet tagging in
this case where squark and gluino masses are comparable. This is in
contrast to the gain in reach of $\sim 8$\% for the case of heavier
squarks in the right hand frame.

\section{Top tagging and the reach of the LHC}\label{sec:toptag}

We have seen that requiring a $b$-tagged jet reduces the SM background
relative to the SUSY signal in a wide variety of models, and so
increases the SUSY reach of the LHC. This then raises the question
whether it is possible to further increase this reach by requiring a
top-tagged jet, since the mechanisms that serve to enhance the decays of
SUSY particles to $b$-quarks frequently tend to enhance decays to the
entire third generation of SM fermions. SM backgrounds to $\eslt$ events
with $t$-quarks should, of course, be smaller than those for events with
$b$-quarks. In this section, we study the prospects for top tagging,
once again using the inverted mass hierarchy model line (\ref{imhneg})
to guide our thinking.

Top tagging in SUSY events has been suggested previously for the
reconstruction of SUSY events, assuming that $\tst_1$ or $\tb_1$ are
light enough so that $\tg \to t\tst_1 \to t b\tw_1$ and/or $b\tb_1 \to b
t\tw_1$ occur with large branching fractions \cite{nojiri}. Using a top
reconstruction procedure described below, together with an estimate of
fake tops from an analysis of side-bands,  
it was shown that for
$m_{\tg}\sim 700$~GeV, for which the SUSY event rate is very large,
partial reconstruction of SUSY events with gluinos decaying to third
generation squarks was possible at the LHC.

We follow the approach developed in this study to reconstruct the top
quark via its hadronic decay mode. In a sample of multi-jet +
$\eslt$ events with at least one tagged $b$-jet, we identified a
hadronically decaying top by first identifying all pairs of jets
(constructed from those jets that are not tagged as a $b$-jets) as a
hadronically decaying $W$ if $|m_{jj}-M_W| \leq 15$~GeV. We then
pair each such $W$ with the tagged $b$-jet(s) and identify any
combination as a top if $|m_{bW}-m_t|\leq 30$~GeV. If we can reconstruct
such a ``top'', we defined the event to be a top-tagged event.  The
efficiency for tagging tops in this way turns out to be small.\footnote{
In a simulated sample of about 90K $t\bar{t}$ pairs with a hard
scattering $E_T$ between 50--400~GeV, we found only 6,255 top tags even
with $\epsilon_b=1$. To understand this large loss of efficiency we
note that first, leptonically decaying tops (branching
fraction of $\sim 1/3$) are clearly not identified. Second, $b$-jets are
within their fiducial region ($E_{Tj}>40$~GeV, $|\eta_j|\leq 1.5$,
with a $B$-hadron with $p_T(B)\geq 15$~GeV within a cone of $\Delta
R=0.5$ of the jet axis) only about 5/8 of the time. Third, it is
necessary for the top with the $b$-jet inside the fiducial region to
decay hadronically in order to make the top mass window, since the wrong
combination mostly falls outside. Finally, if the jets from the $W$ from
the top with the tagged $b$ merge or radiate a separate jet at a large
angle, this $W$ is lost, and hence the top, is not tagged. We have
checked with our synthetic top sample that the choice of mass bins of
$\pm 15$~GeV about $M_W$ and $\pm 30$~GeV about $m_t$ suggested in
Ref.~\cite{nojiri} does not lead to loss of signal from events where the
top decays hadronically into well separated jets: most of the loss in
efficiency comes from the other factors detailed above.}

For our examination of the impact of top tagging on the SUSY reach of
the LHC, we have again chosen the $SO(10)$ model line (\ref{imhneg})
with $\mu < 0$ as a test case. In this case, since other squarks are
heavy, the gluino mainly decays with roughly equal likelihood via $\tg
\to \tst_1 t$ and $\tg \to \tb_1 b$, where subsequent decays of the
third generation squarks can lead to yet more top quarks in SUSY
events. As for the case of $b$-jet tagging, we have run the SUSY sample
through a set of cuts shown in Table~\ref{tab:topcuts} to optimize our
top-tagged signal relative to SM background. Because of the small
efficiency for top-tagging we cannot, however, afford a large reduction
of the signal from multiple cuts. We have, therefore, restricted our
optimization to cuts on just the three variables $\eslt$, $m_{\rm eff}$
and $n_j$, imposing the basic requirements on the observability of the
signal discussed in Sec.~\ref{sim}.
%%% 
\begin{table}[htb]
\begin{center}
\begin{tabular}{lcc}
\hline \hline
Variable & Values \\
\hline
$\eslt ({\rm GeV}) \ge$ & $300, 400,..., 900$ \\
$m_{\rm eff} ({\rm GeV}) \ge$ & $800, 900,..., 2000$ \\
$n_j\ge$ & $3, 4,..., 8$ \\
$n_b\ge$  &   $1$ \\
\hline \hline
\end{tabular}
\end{center}
\caption{The complete set of cuts examined for extraction of the
  SUSY signal with tagged $t$-jets. In addition to the basic cuts detailed
  in the text, we require that $S_T \ge 0.1$.  }
\label{tab:topcuts}
\end{table}

The results of our SUSY reach analysis with
top-tagging are summarized in Table~\ref{tab:toptag}. Here, we show the
optimized statistical significance of the SUSY signal for three cases
in the vicinity of the ultimate reach using this
technique. In this table, we show representative sparticle masses along
with branching fractions for sparticle decays that lead to top quark
production in SUSY cascades. We then detail the final choice of cuts
that optimizes the top-tagged SUSY signal. We also show the top-tagged
signal cross section after these cuts along with the corresponding SM
background, and the statistical significance of the top-tagged signal
achieved in cases 1 and 2; for case~3, the signal is not observable by
our criteria. Finally, in the last two rows we show the corresponding
statistical significance using $b$-jet tagging discussed in
Sec.~\ref{sec:btag}. We see from the Table that while top tagging allows
an LHC reach for $m_{\tg}$ just above 1600~GeV, {\it the top-tagged  rate
becomes too low} for heavier gluinos. In contrast, $b$-jet tagging
yields a statistical significance in excess of 50 close to the
top-tagged reach. We thus conclude that while top-tagging can be used as
a diagnostic tool, or even for reconstruction of SUSY events
\cite{nojiri} in favourable cases, it will not extend the SUSY reach of
the LHC. 

\begin{table}[htb]
\begin{center}
\begin{tabular}{lcccc}
\hline \hline
& CASE 1  & CASE 2  & CASE 3  \\
\hline
$m_{16}$~(GeV) & 1650 & 1770   & 1820  \\
%$m_{1/2} $~(GeV) & 642 & 685  & 703  \\
%$A_{0}$~(GeV) & -3300 & -3540 & -3640 \\
%tan$\beta$ &47 &47 & 47\\
%sign($\mu$) & -1 & -1& -1\\ \hline  
$m_{\tg}$ (GeV)  & 1522 & 1614  & 1661 \\
$m_{\tu_R}$~(GeV) & 2108 & 2255 & 2319 \\
$m_{\tst_1}$~(GeV) & 714 & 766 & 792 \\
$m_{\tb_1}$~(GeV) & 744 & 842 &  876 \\
$m_{\tw_1}$~(GeV) & 533 & 570 &589 \\
$m_{\tz_1}$~(GeV) & 279 & 299 & 309 \\
$B(\tst_1 \to t\tz_i)$ & 0.64 & 0.69 & 0.70 \\
$B(\tb_1 \to t\tw_1)$ & 0.37 & 0.31 & 0.30 \\ \hline
$\eslt $ (GeV) $\ge$ &  300 & 500 & n/a \\
$m_{\rm eff}$ (GeV $\ge$ & 1700 & 800 & n/a \\
$n_j\ge$ & 8  &  3  & n/a \\ \hline
$\sigma_{\rm SUSY}$ (fb) & 0.138 & 0.108 &  n/a  \\  
$\sigma_{\rm back}$ (fb) & 0.0117 & 0.0306 &  n/a \\ \hline
$N_{\rm SUSY}/\sqrt{N_{\rm back}}$ &  & & \smallskip\\
top tag  & 12.7 & 6.14 & 0.00 \\
$1b$ & 62.8 & 52.5 & 44.7  \\
$2b$ & 93.5 & 64.0 & 46.4 \\
\hline \hline
\end{tabular}
\end{center}
\caption{A comparison of the statistical significance of the LHC signal
using top-tagging described in the text,  for three
different cases along the $SO(10)$ model line (\ref{imhneg}), with
other parameters as fixed by Eq.~(\ref{above}).  The first few lines
show the value of $m_{16}$ along with sample particle masses and
branching fractions. The next three lines show the choice of cuts for
the variables in Table~\ref{tab:topcuts} that maximizes the statistical
significance of the top-tagged signal. The signal and SM background
cross sections for these cuts are shown on the next two lines for the
cut choice that leads to an observable signal with the greatest
statistical significance. The last
three rows compare the statistical significance of the signal using
top-tagging with that obtained using $b$-jet tagging discussed in
Sec.~\ref{sec:btag}.}
\label{tab:toptag}
\end{table}

\section{Can we directly detect third generation squarks?} \label{third}

In models where the third generation is significantly lighter than the
other generations, it is natural to ask whether it is possible to detect
signals from the {\it direct production} of third generation
squarks. As already mentioned, their detection as secondaries from
production and subsequent decays of gluinos is possible if the gluino
itself is not very heavy \cite{nojiri}. Our goal, therefore, is to
examine whether the signal from the direct production of third
generation squarks can be separated both from SM backgrounds, as well as
from production of other SUSY particles. Clearly, this is a
model-dependent question, since the SUSY ``contamination'' to the third
generation signal will depend strongly on the masses of the other
squarks and the gluino. In this section, we will study this issue within
the context of the inverted mass hierarchy model with $\mu < 0$, that
we have used as our canonical test case. 

Since there are essentially no third generation quarks in the proton,
the cross section for third generation squarks falls rapidly with the
squark mass, and the signal becomes rapidly rate-limited. Therefore, we
confine ourselves to the signal from third generation squarks with
masses around 300--500~GeV, where the signal is likely to be the
largest. To unequivocally separate out the third generation signal, we
must use cuts that are hard enough to reduce the SM backgrounds to
acceptable levels, yet not so hard as to enhance the ``contamination''
from heavier sparticles that, though they are produced with (much) smaller
cross sections than third generation squarks, would pass these hard cuts
with much larger efficiency.

Since third generation sfermions decay preferentially to third
generation fermions (we focus on the case where $\tst_1 \to b\tw_1$ is
accessible), we study the signal with at least one tagged $b$-jet.  We
found, however, that even the softest set of cuts in
Table~\ref{tab:cuts} that we actually use for our analysis of the
SUSY $b$-tagged signal, are too hard for the purpose of extracting the signal
from third generation squarks. We, therefore, returned to our basic
cuts, 
$$ \eslt > 100~{\rm GeV}, E_T(j_1, j_2)> 100~{\rm GeV}$$
and augmented these with the requirements,
$$E_T(j_3, j_4) > 100~{\rm GeV}, \ S_T \ge 0.1, \ n_b\ge 1, $$ and ran
the third generation signal through the analysis cuts in Table~\ref{tab:3rdgen}
to extract the optimal $N_{\rm signal}/N_{\rm back}$ ratio 
(where the background includes the SM
and the SUSY contamination as we discussed).  These cuts, which are
applied ``from below'', primarily serve to control the SM background
which is very large after just the basic cuts (see Table~\ref{tab:bkg}),
but reduced by the additional requirements of a tagged $b$-jet and two
additional 100~GeV jets.

%Table of 3rd generation loops 
\begin{table*}[htb]
\begin{center}
\begin{tabular}{lc}
\hline \hline
Variable & Values \\
\hline
$\eslt$ (GeV) $\ge$ & $100, 150, 200, 250$ \\
$[E_T(j_1),E_T(j_2)] $ (GeV) $\ge$ & $(100, 100), (200, 100), (200, 150),
(300, 100), (300, 150), $ \\
& $(300, 200), (400, 100), (400, 150), (400, 200)$ \\
$E_T(j_3)$, $E_T(j_4)$ (GeV) $\ge$ & 100 \\
$E_T(b_1)$ (GeV) $\ge$   & $40, 100, 200, 300, 400$ \\
$m_{\rm eff}$~(GeV) $\ge$ & $500, 600, ..., 1500$ \\
$n_j\ge$ & $4, 5, 6, 7 $\\
 $n_b\ge$  &   $1$ \\
$S_T \ge$ & 0.1 \\
\hline \hline
\end{tabular}
\end{center}
\caption{The set of cuts imposed from below 
 that we examined for our study of the
  extraction of the third generation squark signal at the LHC. 
 Additional cuts were also imposed from
  above to optimize the third generation squark signal relative to that
  from the production of heavier gluinos and squarks of the first two
  generations, as discussed in the text and detailed in
  Table~\ref{tab:thirdgenfinal}.}
  
\label{tab:3rdgen}
\end{table*}

We show the results of our analysis in Table~\ref{tab:thirdgenfinal}. 
The parameters are shown in the
first four rows of the Table, while the next few rows show
representative sparticle masses. The first two cases are along the $\mu<
0$ model line that we had introduced previously. In the first two cases
$B(\tst_1\to b\tw_1)=1$, while in Case~3, $B(\tst_1\to b\tw_1)= 0.74$,
with the remainder being made up by the decay $\tst_1\to t\tz_1$. The
next several rows list the optimized choice of cuts from the $4\times
9\times 5\times 11\times 4$ possibilities in Table~\ref{tab:3rdgen}, along with
the cross sections for ({\it i})~the third generation signal, ({\it
ii})~the SM background, and ({\it iii})~the ``SUSY contamination''
defined as the SUSY signal from production of sparticles other than
third generation squarks, after these cuts. We see from these cross
sections that both the event rates and the statistical significance of
the third generation signal (even with the SUSY contamination included
in the background) is very large. The problem, however, is that the signal
to background ratio is smaller than 0.1, if the SUSY contamination is
included in the background, and fails to satisfy our observability
criterion.\footnote{Many authors do not impose such a requirement on the
observability of the signal. We believe that some requirement on the
$N_{\rm signal}/N_{\rm back}$
 ratio is necessary since otherwise a signal with 5K events, above
a background of 1M events would be considered significant. This would 
indeed be the case if the background were known to a very high
precision; however a systematic uncertainty of 0.5\% on the background
could clearly wipe out the signal, at least if the signal is extracted
by subtracting the theoretically calculated background! In the case at
hand, where the SUSY model is not {\it a priori} known, and has to be
arrived at using the same data, it is clear that subtraction of the SUSY
contamination will suffer from considerable uncertainty until the data
and theory both become mature enough for such a subtraction to be
carried out. While our criterion requiring $N_{\rm signal}/N_{\rm back}>0.25$ is admittedly
arbitrary, we believe that it is necessary to impose some lower limit on
the signal to background ratio for a semi-realistic assessment.} We can,
however, reduce the SUSY contamination (primarily from heavier
sparticles) relative to the third generation signal by requiring that
{\it the signal is not too hard.} Toward this end, we impose an {\it
upper limit}, $m_{\rm eff} < 1000$~GeV, which efficiently reduces the
contamination from heavy sparticles with correspondingly modest
reduction of the cross sections from the softer third generation and SM
processes. The corresponding cross sections after this cut are shown on
the next three rows of the Table, while the last row shows the final two
signal to total background ratio that we are able to obtain, along with
the statistical significance of the third generation signal with an
integrated luminosity of 100~fb$^{-1}$. 

%\newpage
%
\begin{table*}[htb]
\begin{center}
\begin{tabular}{lcccc}
\hline\hline
 &CASE 1 & CASE 2  & CASE 3  \\
\hline
$m_{16}$ (GeV) &717 & 854   & 739  \\
$m_{1/2} $ (GeV) & 306 & 355  & 361  \\
$A_{0}$  (GeV) & -1434 & -1708 & -1478 \\
tan$\beta$&47 &47 & 47\\ \hline
$\mu$ (GeV) & -372 & -428& -477\\
$m_{\tg}$ (GeV)  & 764 & 879  & 886 \\
$m_{\tu_R}$ (GeV) & 966 & 1127 & 1070 \\
$m_{\tst_1}$ (GeV) & 274 & 316 & 460 \\
$m_{\tb_1}$ (GeV)& 442 & 559 &  400 \\
$m_{\tw_1}$ (GeV) & 236   & 279 & 287 \\ \hline
$\eslt$ (GeV) $>$ & 150 & 100 & 150 \\ 
$[E_T(j_1), E_T(j_2)]$ (GeV) $>$ & (100,100) & (100,100) & (200,100) \\ 
$E_T(b_1)$ (GeV) $>$ & 40 & 40 & 40 \\
$m_{\rm eff}$ (GeV) $>$& 500 & 500 & 600 \\ 
$n_j\ge$ & 5 & 6 & 4 \\
$\sigma_{{\rm 3rd \ gen.}} $ (fb) & 120.2 & 74.1 & 80.6 \\ 
$\sigma_{\rm SUSY \ cont.} $ (fb) & 1176.3 & 590.6 & 828.9 \\ 
$\sigma_{\rm SM}$ (fb) & 432.6 & 454.1 & 580.4 \\ \hline 
$m_{{\rm eff}}$ (GeV) $<$  & 1000 & 
1000 & 1000 \\ 
$ \sigma_{{\rm 3rd \ gen.}} $ (fb) & 47.2 & 30.9 & 20.5 \\
$\sigma_{\rm SUSY \ cont.}$ (fb) & 109.5 & 42.0 & 40.0\\ 
$\sigma_{\rm SM}$ (fb) & 141.7 & 180.6 & 155.1\\
$\sigma_{{\rm 3rd \ gen.}}/\sigma_{\rm tot. \ bkg}$ & 0.188 & 0.14 & 0.105\\
$N_{\rm signal}/\sqrt{N_{\rm back}}$ & 29.8 & 20.7 & 14.7 \\
%\\ add cut & $nj < 9$ &$ nj < 9$
%& $nj < 8 $, $ pj1 < 300$ \\ $N_{signal}/N_{totbg}$ & 0.186(45.6) &
%0.142(29.6) & 0.117(217.3) \smallskip\\ $N_{signal}/N_{susynet}$ &
%0.44(149.7) & 0.88(63.8) & 0.51(51.3)\\ $N_{signal}/N_{stdmod}$ &
%0.32(141.4) & 0.17(174.3) & 0.15(114.1)\\ decay & $\tw_{1}$b 1.000 &
%$\tw_{1}$b 1.000 & $\tz_{1}$t 0.263 , $\tw_{1}$b 0.737 \\ 
\hline\hline
\end{tabular}
\end{center}
\caption{Optimized cuts, along with cross sections for the signal from
  direct production of light third generation squarks, for SM
  background, and for SUSY contamination to the third generation squark
  signal.
%   The first four rows
%   specify the input parameters for our three case studies while the next
%   six rows specify $\mu$ and selected sparticle masses.  
  Input parameters and selected sparticle masses are shown in the first
  ten rows.
  The next
  several rows detail our choice of cuts from the set in
  Table~\ref{tab:3rdgen}, %chosen to ameliorate the softer Standard Model
 % background, 
along with cross sections for the third generation signal,
  for SUSY contamination,
% to this signal from other SUSY sparticles, 
  and for the
  SM background after these cuts. The last six rows
  show the cut ``from above'' discussed in the text, along with our
  final results for the observability of the third generation signal
  over total backgrounds, including SUSY contamination.} 
 %s for the various cross sections, the signal to total background
 % ratio (including SUSY contamination) and the statistical significance
 % of the signal.}
\label{tab:thirdgenfinal}
\end{table*}

Several comments about the Table are worth noting. 

\begin{itemize}
\item We see from the Table that before the cut restricting the value of
  $m_{\rm eff}$ from above, the background was dominated by SUSY
  contamination. In contrast, after this cut, the dominant source to the
  background comes from SM processes.

\item With the cuts that we have devised, the event rates for the third
  generation signal as well as its statistical significance are
  large. For reasons already discussed, we do not, however, believe that
  it will be easy to unequivocally ascertain the direct production of
  third generation squarks in the signal. For this to be unambiguously
  possible, it will be necessary to have an understanding of the
  contributions from other SUSY sources to the event rate after our
  cuts. This may well be possible because with hard cuts it should be
  possible to isolate the signal from heavy squarks and gluinos where
  contamination from both SM and the lighter third generation squarks is
  small. Just how well it will be possible to extrapolate this measured
  signal into ``softer kinematic regions'' will determine the precision
  with which the SUSY contamination can be subtracted. This issue is
  beyond the scope of the present analysis. 

%\medskip\\
%
%\clearpage

\item We examined additional cuts on $E_T(j_1, j_2)$ and $n_j$ to see if
  we could raise the signal to background ratio. We found that a small
  increase ($\sim 10$\%) may indeed be possible by restricting $n_j$
  from above to be smaller than 8 or 9. Since our calculation of the
  background with high jet multiplicity is carried out only in the
  shower approximation, we did not feel that our estimate of this
  improvement is reliable, and choose not to include it in the Table.

\item We stress again that the SUSY contamination is model-dependent. We
  can see from the Table that if gluinos and other squarks are indeed
  decoupled at the LHC, and only third generation squarks are light,
  their signal should be readily observable in all three cases. 

\end{itemize}

\section{Charm-jet tagging} \label{ctag}

Charm jet tagging offers a different possibility for enhancing the SUSY
signal, especially in the case where a light top squark dominantly
decays via
$\tst_1 \to c\tz_1$. Charm jets may be tagged via the detection of a
soft muon within the jet. Muons inside jets also arise from
semi-leptonic decays of $b$-quarks and from accidental
overlaps of unrelated muons with jets. Since $m_b$ is significantly
larger than $m_c$, the variables $|{\vec{p}}_T^{\; \rm rel}|\equiv
|{\vec{p}}_T(\mu)\times{\hat{p}}_j|$ and $\Delta R(\mu,j)\equiv
\sqrt{\Delta\phi(\mu,j)^2+\Delta\eta(\mu,j)^2}$ can serve to distinguish
muon-tagged $c$-jets from correspondingly tagged $b$-jets or accidental
overlap of an unrelated muon with jets. Charm jet tagging with soft
muons was first examined in Ref.~\cite{bst} as a way of enhancing the
$t$-squark signal from $p\bar{p} \to \tst\tst X \to cc+\eslt +X$
production at Run I of the Fermilab Tevatron, but was found to have a 
reach smaller than the reach obtained via the conventional $\eslt$
analysis  {\it because the muon-tagged signal was severely
  rate-limited}. It was, however, subsequently shown 
that using soft muons to tag the $c$-jet indeed enhances the
top squark reach \cite{sender} but only  for an integrated
luminosity larger than $\sim$~1 fb$^{-1}$, available today after the
upgrade of the Main Injector. 

These considerations led us to examine whether charm tagging may be
similarly used at the LHC, at least for the case where $\tst \to
c\tz_1$. Since the goal is to separate the charm jets from the decay of
$\tst_1$ from other SUSY sources (which are frequently rich in
$b$-jets), it is crucial to be able to separate the $c$ and $b$ jets
with at least moderate efficiency and purity.  Following
Ref.~\cite{sender}, we examined many strategies to obtain this separation
in the plane formed by the variables $|{\vec{p}}_T^{\; \rm rel}|$ and $\Delta
R(\mu,j)$ but without any success. The difference between the situation
at the Fermilab Tevatron, where this strategy appears to be moderately
successful, and the LHC is the kinematics of the events. In contrast to
the Tevatron, where jets with $E_T > 25$~GeV are readily detectable, at
the LHC we have required $E_T(j)>50$~GeV in order not to be overwhelmed
by mini-jet production. For this harder jet kinematics, the difference
between $m_b$ and $m_c$ appears to be too small to yield significant
separation between $c$- and $b$-jets that are not vertex-tagged. The
larger contamination from $b$-jets at the LHC only exacerbates this
situation.

Before closing this section, we also mention one other (also
unsuccessful) strategy that we tried for $c$-tagging. The idea was to
utilize the difference in the distributions of $z\equiv E_{\mu}/E_c$ for
muons of a fixed sign of the charge from $b$ or $\bar{c}$ decays. While
the expected distributions from the quark decays are indeed significantly
different, this strategy also fails because these quarks hadronize
before they decay, and the $z$-distributions of the muons from the
corresponding bottom or charm meson decays are essentially the same. 

\section{Summary} \label{summary}

The search for gluinos and squarks of supersymmetry is an important item
on the agenda of LHC experiments. In most models $m_{\tg} \lesssim m_{\tq}$,
so that, except when squarks and gluinos are close in mass, we expect
squarks to decay mainly to gluinos. The decay patterns of the gluino
will then determine the topologies of the bulk of SUSY events at the
LHC. Generally speaking, we expect that sparticle production at the LHC
will be signalled by an excess of $n$-jet + $m$-isolated-leptons +
$\eslt$ events (possibly together with isolated hard photons), with the
relative rates for the various topologies determined by sparticle decay
patterns. However, there are a
number of well-motivated models (see Sec.~\ref{models}) where gluinos
preferentially decay to third generation fermions, so that SUSY events
are likely to include $b$- or even $t$-jets. Since a large part of the
SM background to the inclusive jets + $\eslt$ SUSY signal comes from
$V+\eslt$ production ($V=W,Z$) and from QCD, $b$-jet tagging should
serve to discriminate between the SM and SUSY sources of missing
transverse energy events at the LHC.  In the HB/FP region of the mSUGRA
model favoured by the WMAP determination of the relic density,
$b$-tagging increases the LHC reach for gluinos by $\sim 20$\%
\cite{MMT,india} depending on the tagging efficiency and purity that is
ultimately attained.  Though this does not appear to be a large
enhancement, we must remember that we are probing the gluino mass range
where the production cross section is already small due to kinematic
considerations.

In Sec.~\ref{sec:btag}, we have examined the impact of $b$-tagging on
the LHC reach for a variety of models introduced in
Sec.~\ref{models}. We use a conservative projection for the tagging
efficiency of $b$-jets in the high luminosity LHC environment: 50\% for
central $b$-jets with $E_T \ge 40$~GeV. We find that while $b$-tagging
does indeed increase the SUSY reach of the LHC, the enhancement is
typically {\it smaller} than that found for the HB/FP region of the
mSUGRA model. In this model squarks are in the multi-TeV range,
consequently, the SUSY
signal (after selection cuts) comes mainly from the pair production of
gluinos, whose decays are ``$b$-rich'' as we mentioned above. This same
enhancement is not obtained in models where squarks and gluinos have
comparable masses. Then, the branching fraction for the decay $\tq \to
q\tg$ is kinematically suppressed, and the squarks mainly decay via $\tq
\to q\tz_i$ and $\tq \to q'\tw_i$, where the daughter quark (mostly)
belongs to the same generation as the parent squark. Since (for values
of $m_{\tq} \gtrsim 1$~TeV) first generation squarks are much more
abundantly produced at the LHC than squarks of other generations by the
collisions of (first generation) valence quarks in the proton, their
decays do not lead to $b$-jets. As a result, (and this is confirmed by
our results in Sec.~\ref{sec:btag}) $b$-tagging enhances the LHC
reach the most in models where squarks are significantly heavier than
gluinos: if instead squark production is the origin of a substantial
portion of the SUSY $\eslt$ signal, $b$-tagging will not be helpful, and
could even lead to a degradation of the reach of the LHC.

Since the mechanisms that lead to enhanced decays of gluinos to
$b$-quarks mostly revolve around the large third generation Yukawa
couplings, these typically also enhance sparticle decays to $t$-quarks
(if these are not kinematically suppressed). This led us to examine in
Sec.~\ref{sec:toptag} whether tagging $t$-jets (which potentially
reduces SM backgrounds even more efficiently than $b$-jet tagging does)
could lead to an increased reach for SUSY. We found, however, that this
is not the case because the top-tagging efficiency is very
low. Top-tagging may, however, facilitate the reconstruction of SUSY
events in favourable cases \cite{nojiri}, and furthermore, can be used
to confirm a SUSY signal first detected in other channels.

There are well-motivated models with an inverted squark mass hierarachy,
where third generation squarks are much lighter than other squarks and
gluinos. These models would be strikingly confirmed if signals from 
the direct production of these relatively light top and sbottom squarks
as well as from their production as secondaries from gluino decays
could be separately identified. Since the latter possibility has already
been studied in Ref.~\cite{nojiri}, we concentrated on the signal from direct 
production of third generation squark pairs in Sec.~\ref{third}. We found
that while these may be readily separated from SM backgrounds, it may be
more difficult to discriminate between them and the signal from the
production of heavier gluinos and first generation squarks. This is,
however, clearly a model-dependent question. It could be that gluinos and first
generation squarks are so heavy that the SUSY contamination is not an
issue at all. Alternatively, if gluinos and first generation squarks are
not extremely heavy, it may be possible 
to determine their properties  by studying the event sample with
very hard cuts to remove contributions from the production of the much
lighter third generation squarks; we can  then use these to subtract
the 
contamination from gluino and first generation  squark production
in the analysis of the signal from the direct production of third
generation squarks.

Finally, in Sec.~\ref{ctag} we have examined whether charm-tagging
(using muons inside a jet) may be useful to enhance the SUSY signal at
the LHC. Our conclusions are pessimistic. The kinematics of events at
the LHC (in contrast to the kinematics at the Fermilab Tevatron) makes
it very difficult to distingish between $b$-jets and $c$-jets using the
soft-muon-tagging technique.

In summary, we have found that the use of $b$-tagging enhances the SUSY
reach of the LHC by up to 20\% in a variety of well-motivated models,
with the largest increase in reach being obtained in models with
$m_{\tq} \gg m_{\tg}$, and no increase (or even a reduction in reach) if
squark production is an important part of the signal after the final
cuts.
For
many, but not all, models tagging $b$- and $t$-quark jets improves the
signal to background ratio, resulting in a cleaner SUSY event sample,
which can then be used either to reconstruct SUSY event chains
\cite{nojiri}, or as has been recently suggested, possibly to determine
the gluino mass \cite{howiefp}.\footnote{The technique suggested here
depends crucially on the ability to reliably calculate the absolute
cross section of the SUSY signal after hard cuts and background
subtraction.}  In contrast, we find that charm-jet tagging does not
appear to be useful at the LHC.

\acknowledgments
This research was supported in part by a grant from the United States
Department of Energy and by Funda\c{c}\~{a}o de Amparo 
\`a  Pesquisa do Estado de S\~ao Paulo (FAPESP). J.~K.~Mizukoshi
would like to thank Instituto de F\'{i}sica da Universidade de S\~ao Paulo
for the use of its facility.

%%%%%%%%%%%%%%%%%%%%%%%%%%%%%%%%%%%%%%%%%%%%%%%%%%%%%%

\end{document}